\documentclass[lettersize,journal,twoside]{IEEEtran}
\usepackage{amsmath,amsfonts}
\usepackage[ruled,linesnumbered]{algorithm2e}
\usepackage{array}
\usepackage{multirow}
\usepackage{textcomp}
\usepackage{stfloats}
\usepackage{url}
\usepackage{verbatim}
\usepackage{graphicx}
\usepackage{cite}
\usepackage{color}
\usepackage{subfigure}
\usepackage{bbding}

\usepackage{footnote}

\usepackage{dsfont}
\usepackage{amsfonts,amsmath}
\usepackage{url,cite}
\usepackage{color}
\usepackage{booktabs}
\usepackage{extarrows}
\usepackage[colorlinks,
linkcolor=blue,
anchorcolor=blue,
citecolor=green]{hyperref}

\begin{document}

\title{Hyperspectral and Multispectral Image Fusion Using the Conditional Denoising Diffusion Probabilistic Model}

\author{Shuaikai~Shi,~\IEEEmembership{Student Member,~IEEE,}
	Lijun Zhang,
		Jie~Chen,~\IEEEmembership{Senior Member,~IEEE} \\
	\thanks{Shuaikai Shi, Lijun Zhang and Jie Chen are with the Center of Intelligent
		Acoustics and Immersive Communications, School of Marine Science and
		Technology, Northwestern Polytechinical University, Xi’an 710072, China,
		and also with the Key Laboratory of Ocean Acoustics and Sensing, Ministry of Industry and Information Technology, Xi’an 710072, China (e-mail:\_shuaikai\_shi@mail.nwpu.edu.cn; zhanglj7385@nwpu.edu.cn; dr.jie.chen@ieee.org).}
}

\markboth{Submission to IEEE TRANSACTIONS ON GEOSCIENCE AND REMOTE SENSING,~Vol.~XX, No.~XX, ~2023}%
{Shi \MakeLowercase{\textit{et al.}}:   Hyperspectral and Multispectral Image Fusion Using the Conditional Denoising Diffusion Probabiblistic Model}


\maketitle

\begin{abstract}
Hyperspectral images (HSI) have a large amount of spectral information reflecting the characteristics of matter, while their spatial resolution is low due to the limitations of imaging technology.
Complementary to this are multispectral images (MSI), e.g., RGB images, with high spatial resolution but insufficient spectral bands.
Hyperspectral and multispectral image fusion is a technique for acquiring ideal images that have both high spatial and high spectral resolution cost-effectively.
Many existing HSI and MSI fusion algorithms rely on known imaging degradation models, which are often not available in practice.
In this paper, we propose a deep fusion method based on the conditional denoising diffusion probabilistic model, called DDPM-Fus.
Specifically, the DDPM-Fus contains the forward diffusion process which gradually adds Gaussian noise to the high spatial resolution HSI (HrHSI) and another reverse denoising process which learns to predict the desired HrHSI from its noisy version conditioning on the corresponding high spatial resolution MSI (HrMSI) and low spatial resolution HSI (LrHSI).
Once the training is completes, the proposed DDPM-Fus implements the reverse process on the test HrMSI and LrHSI to generate the fused HrHSI.
Experiments conducted on one indoor and two remote sensing datasets show  
the superiority of the proposed model 
when compared with other advanced deep learning-based fusion methods.
The codes of this work will be open-sourced at this address: \url{https://github.com/shuaikaishi/DDPMFus} for reproducibility.

\end{abstract}


\section{Introduction}
\label{sec.sec1}
\IEEEPARstart{S}{pectral} imaging allows for the simultaneous capture of both spatial and spectral information of a scene, providing light reflectance information beyond human perception.
Benefiting from abundant spectral features, this technique has been used in a wide-range of applications, including face recognition \cite{face_recognition}, object detection\cite{object}, remote sensing \cite{remote}, agriculture\cite{agriculture}, etc.
However, the HSI typically has lower spatial resolution than the RGB images due to the larger instantaneous field of view (IFOV) \cite{ifov}.
In practice, it is only possible to capture either HrMSI or LrHSI in a single imaging shot.
This has a significant impact on the accuracy and reliability of the analysis results obtained from the captured data.
Fortunately, HSI-MSI fusion \cite{fusionReview} can fuse a pair of degraded HrMSI and LrHSI to produce the desired HrHSI.
This fusion technique overcomes the limitations of acquiring HrHSI in a single shot and enables a wider range of applications.

\subsection{Motivation}
Classical HSI-MSI fusion approaches generally assume the parameters of the observation model, i.e., point spread function (PSF) in the spatial degradation and spectral response function (SRF) in the spectral degradation, are known. 
However, in practice, the degradation processes are complex and the parameters may be difficult or even impossible to accurately determine \cite{wang2010Improved}.
Therefore, the performance of such conventional models is limited when the degradation models mismatch the actual system.

Deep learning-based methods have already been introduced to
address the HSI-MSI fusion problem, restore the spatial and spectral details of HrHSI and yield desired results.
These models use deep neural networks, such as convolutional neural networks (CNNs) and Transformer \cite{transformer}, to learn the mapping from (HrMSI, LrHSI) pairs to HrHSI.
Thanks to the extensive expressive ability of neural networks, 
these models can produce high-definition HrHSI.
However, on one hand, the most of existing fusion networks fuse
the input degraded images to the output HrHSI at one step, which can be further improved by incorporating
multiple iterative models.
On the other hand,
the regression-based methods may have restricted practical generalization ability on the new test data
caused by overfitting.
Fortunately, this problem can be alleviated by introducing generative models, such as the recently proposed diffusion model\cite{ddpm}.
Inspired by these two points, we propose a novel multi-stage HSI-MSI fusion model.

\subsection{Methodology Overview and Contributions}
Recently, the denoising diffusion probabilistic model (DDPM) \cite{ddpm} has been attracted 
great attention in the community of deep generative models and used for various 
generative tasks, e.g., image generation\cite{dhariwal2021diffusion} and audio synthesis \cite{chen2020wavegrad}.
Furthermore, the DDPM with extra inputs can be used for the conditional generation, such as image superresolution \cite{sr3}, text-to-image generation \cite{stableDiffusion} and 
image editing \cite{imageEdit}, concluding inpainting, colorization and uncropping. 

The DDPM learns to produce a clean output from its noisy version via multiple denoising
steps.
Specifically, the DDPM contains two processes, namely, 
the forward diffusion process and the reverse denoising process. 
In the training phase, the diffusion process 
adds independent Gaussian noise to the clean sample multiple times 
resulting in the final output tending towards a standard Gaussian distribution.
Then the denoising process, constructed by a deep neural network, learns the reverse mapping from  noisy data to the oringinal clean data.
Once training is complete, the DDPM can generate new samples via implementing the denoising process with Gaussian noise inputs.
Moreover, the conditional DDPM introduce an extra input in the reverse denoising process to guide the model outputs
samples related to this input, e.g., produce a high resolution image corresponding to the blur input rather than output an another unrelated image.
Inspired by the conditional DDPM,
we propose an HSI-MSI fusion model, called DDPM-Fus,
using the HrMSI and LrHSI as conditional inputs.
The proposed DDPM-Fus fuses the HrMSI and LrHSI by a U-net architecture and predicts the added Gaussian noises to recover the spatial details and spectral signatures of the desired HrHSI.

The main contributions of this work are summarized as:

\begin{enumerate}
	\item We propose a novel HSI-MSI fusion model by adapting the denoising diffusion probabilistic model to HSI generation with conditional inputs. 
	Our proposed model outperforms commonly used discriminative models and shows great potential for recovering HrHSI.

	\item 
	The proposed model, DDPM-Fus, produces the fused results by multiple iterative denoising that is distinguished it from the general deep learning-based single-step fusion models.

	\item We demonstrate the fusion on three public hyperspectral data and experiment results show the superiority of the proposed conditional generative model compared to 
	the regression-based fusion approaches.
	
\end{enumerate}
The rest of this paper is organized as follows. 
The related work is reviewed in Section \ref{sec.sec2}. 
Section \ref{sec.sec3} presents the proposed DDPM-Fus.
The fusion experiments conducted on three HSI datasets are demonstrated  in Section \ref{sec.sec4}
to show the effectiveness of DDPM-Fus. 
Section \ref{sec.sec5} concludes this paper and provides a discussion.

\section{Related Work}
\label{sec.sec2}
HSI-MSI fusion has been seen as an effective tool to obtain the desired HrHSI.
In our text, the fusion approaches are divided into two types, namely, conventional unsupervised methods, which often use the observed model as priors, and advanced 
supervised deep learning-based models, which usually rely on ground truth HrHSI data for training.

\subsection{Unsupervised HSI-MSI Fusion}
Pansharpening algorithms \cite{pansharpeningReview} have been extended to the HSI-MSI fusion.
Gram–Schmidt adaptive (GSA) \cite{gsa} is a
representative pansharpening algorithm based on component substitution, which uses Gram–Schmidt transformation to separate the spatial component of the LrHSI that needs to be substituted by the HrMSI.
The generalized Laplacian pyramid-based hyper-sharpening (GLP-HS) \cite{glp-hs} adapts another class of pansharpening methods to HSI-MSI fusion that uses the pyramidal
decompositions to obtain the high spatial resolution details in the scene.
Then the spatial details are injected  into  the LrHSI to obtain the desired HrHSI.
Pansharpening-based fusion methods implement fast and independent of observation model, however, these methods may produce HrHSI with coarse spatial details and spectral distortion due to the simplistic model design.

Subspace representation-based fusion methods 
usually use a spectral dictionary to represent all spectral signatures of the HrHSI and then obtain the corresponding coefficients by optimizing an objective function.
Coupled non-negative matrix factorization (CNMF)\cite{cnmf} embeds the linear mixing model (LMM) into the fusion problem and then endmembers of LrHSI are multiplied by abundances of HrMSI to obtain the fused result.
Some constraints are often used to ensure the sparsity of the coefficients and the smoothness of the desired images.
Hyperspectral superresolution (HySure) \cite{hysure} encourages the smoothness of HrHSI using the total variation (TV) regularizers on each band.
A non-negative structured sparse representation (NSSR) \cite{NSSR} uses the clustering method to promote spectral homogeneity.
The tensor representation is also applied to HSI-MSI fusion.
The coupled sparse
tensor factorization (CSTF) \cite{cstf} directly represents the HSI cube by the Tucker
decomposition and constrains the sparsity of the core tensor.
While subspace representation-based fusion methods can get superior performance over sharpening-based algorithms, the fact that most of these methods use a linear representation that limits their use of them.

An alternative to hand-tuning the regularization parameters is to introduce a neural network to learn prior knowledge in the HrHSI. 
An unsupervised sparse Dirichlet-Net (uSDN) \cite{uSDN}
iteratively learns the shared features of abundances in the LrHSI and HrHSI.
A variational autoencoder-like 
probabilistic generative model (NVPGM) \cite{nvpgm} extends uSDN that
global training of the model parameters.
Guide deep decoder (GDD) \cite{GDD}
learns to produce the 
desired HSI from noise based on the deep image prior \cite{dip}. It degrades HrHSI using an observation model and then minimize the errors between outputs and input images, LrHSI and HrMSI.   
These unsupervised deep fusion models
can produce more accurate results, however, these methods usually assume the observation model is known, resulting in a limitation of their practical use.

\subsection{Supervised HSI-MSI Fusion}
Recently, deep learning-based supervised methods have been introduced to HSI-MSI fusion \cite{fusionReview2}. 
Benefiting from the expressive capacity of neural networks, these models learn the mapping from a number of paired LrHSI and HrMSI to the ground-truth images HrHSI in an end-to-end manner.
Once the training is complete, these fusion models have the generalization ability to fuse other degraded image pairs to the desired HrHSI.
Most deep learning-based methods concentrate on exploring  the expressive network modules to 
extract the high spatial resolution information and precision spectral signatures, and then fuse them into output images.
The deep pansharpening network (PANnet) \cite{pannet} with CNNs and the residual block was proposed to fuse a panchromatic image and a corresponding MSI.
Wang \textit{et al.}\cite{xiuheng2021} introduced a deep prior to HSI super-resolution, automatically learning the spatial-spectral priors in the scene.
Hu \textit{et al.} \cite{HSRnet} designed a deep  convolutional  network with the spatial and spectral attention mechanisms to perform the fusion task, called HSRnet.
In addition to the one-stage fusion methods mentioned above,
multi-stage models are expected to achieve better performance.
Zhang \textit{et al.} \cite{ssrnet} proposed a three-stage network consisting of preliminary fusion, spatial and spectral refinement, namely SSRnet.
Xie \textit{et al.} \cite{mhfnet} proposed a physically meaningful HSI-MSI fusion network (MHF-net) which progressively
recover the HrHSI through $K$-stage networks.
Wang \textit{et al.} \cite{EDBIN} proposed an  iterative  fusion method to estimate 
both the observation model and fusion process, namely enhanced deep blind hyperspectral image fusion network (EDBIN).
Rather than using known PSF and SRF as in the unsupervised methods, the deep supervised fusion models leverage the high capacity of deep neural networks to restore the HrHSI and achieve state-of-the-art performance.
These fusion networks are independent of the observation model, showing promising potential for application in practice.
In summary, the properties of several representative fusion methods are shown in Table \ref{tab.methods}.

\begin{table}[]
		\centering
	\renewcommand\arraystretch{1.2}
	\caption{The properties of representative fusion methods.}
	\label{tab.methods}
	\begin{tabular}{|l|c|c|c|c|}
		\hline
		Category                                            & Method  & Ground truth                        & PSF                                          & SRF                                          \\ \hline
		{Pansharpening}                   & GSA\cite{gsa}     & \multirow{2}{*}{\XSolidBrush} & \multirow{2}{*}{\XSolidBrush} & \multirow{2}{*}{\XSolidBrush} \\ \cline{2-2}
		-based& GLP-HS\cite{glp-hs}  &                                              &                                              &                                              \\ \hline
		\multirow{6}{*}{Subspace-based}                     & HySure\cite{hysure}  & \multirow{4}{*}{\XSolidBrush} & \XSolidBrush                  & \XSolidBrush                  \\ \cline{2-2} \cline{4-5} 
		& CNMF\cite{cnmf}    &                                              & \Checkmark                    & \Checkmark                    \\ \cline{2-2} \cline{4-5} 
		& CSTF  \cite{cstf}  &                                              & \Checkmark                    & \Checkmark                    \\ \cline{2-2} \cline{4-5} 
		& NSSR  \cite{NSSR}  &                                              & \Checkmark                    & \Checkmark                    \\ \hline

		                    & uSDN  \cite{uSDN}  & \multirow{3}{*}{\XSolidBrush} & \XSolidBrush  & \Checkmark  \\ \cline{2-2} \cline{4-5} 
	{Unsupervised   deep }	& NVPGM \cite{nvpgm} &                               &\XSolidBrush   & \Checkmark  \\ \cline{2-2} \cline{4-5} 
	learning-based	        & GDD   \cite{GDD}   &                               & \Checkmark    & \Checkmark  \\ 		\hline

 	\multirow{3}{*} & PANnet\cite{pannet} & \multirow{5}{*}{\Checkmark}   & \multirow{5}{*}{\XSolidBrush} & \multirow{5}{*}{\XSolidBrush} \\ \cline{2-2}
                                 {Supervised   deep }           & HSRnet\cite{HSRnet} &                               &            &                                              \\ \cline{2-2}
                                          learning-based	  & SSRnet\cite{ssrnet} &                               &            &                                              \\ \cline{2-2}
                                          & MHF-net\cite{mhfnet} &                               &            &                                              \\ \cline{2-2}
                                             &EDBIN\cite{EDBIN} &                               &            &                                              \\ \cline{2-2}
                         & DDPM-Fus (ous)  &                          &                          &                                              \\ \hline
	\end{tabular}
\end{table}
\begin{figure}[t]
	\centering
	\includegraphics[width=9cm]{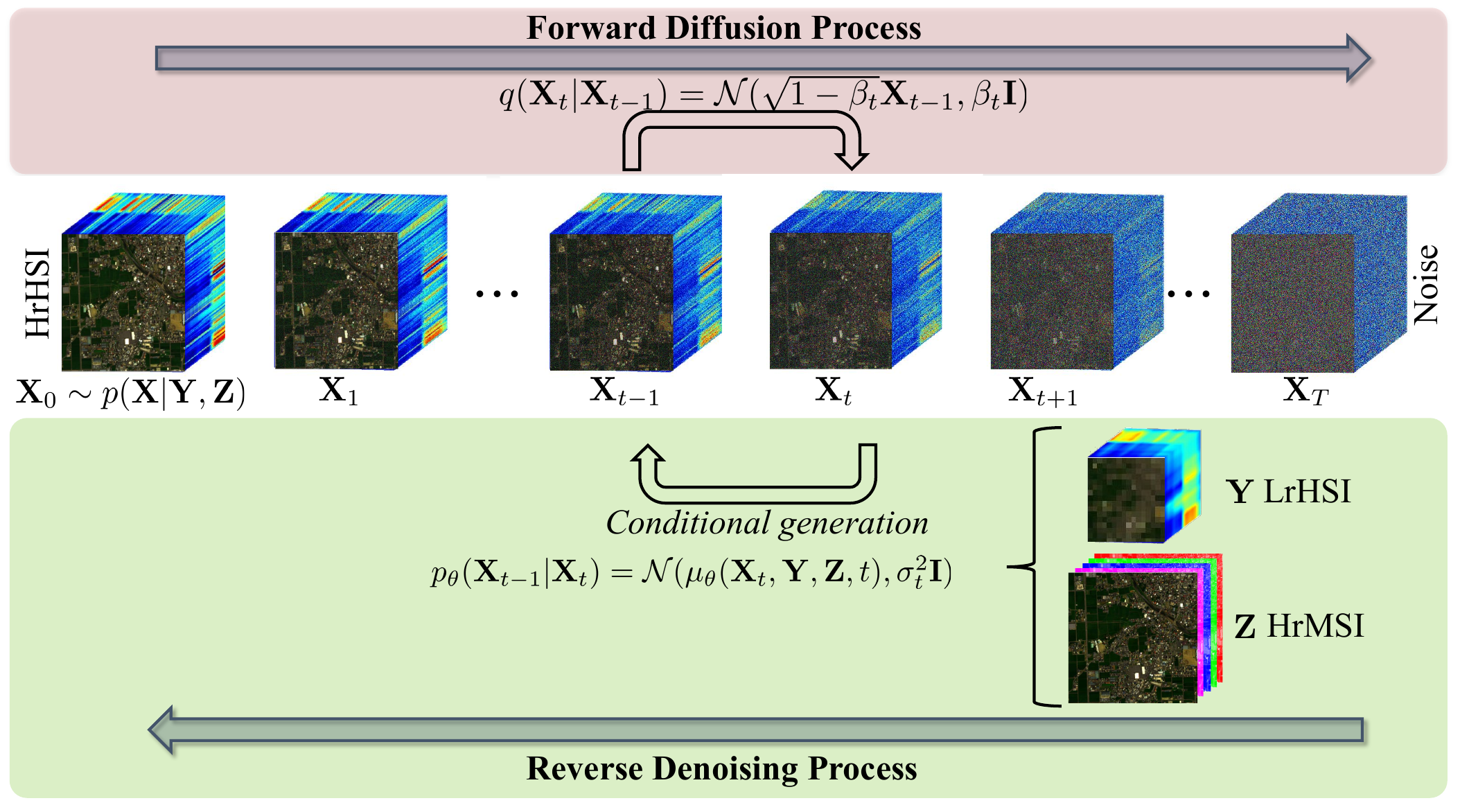}
	\caption{
		Framework of the proposed DDPM-Fus, including the forward process and the reverse process.
		In the forward diffusion process, Gaussian noise is gradually added to HrHSI $\mathbf{X}_0$ over $T$ time steps.
		In the reverse denoising process, the neural network $\mu_\theta(\cdot)$
		learns to recover the  spatial details and spectral signatures of the desired HrHSI by implementing denoising progressively conditioned on two degraded images, $\mathbf{Y}$ and $\mathbf{Z}$.
	}
	\label{fig.ddpm}
\end{figure}

\begin{figure*}[!t]
	\centering
	\includegraphics[width=18cm]{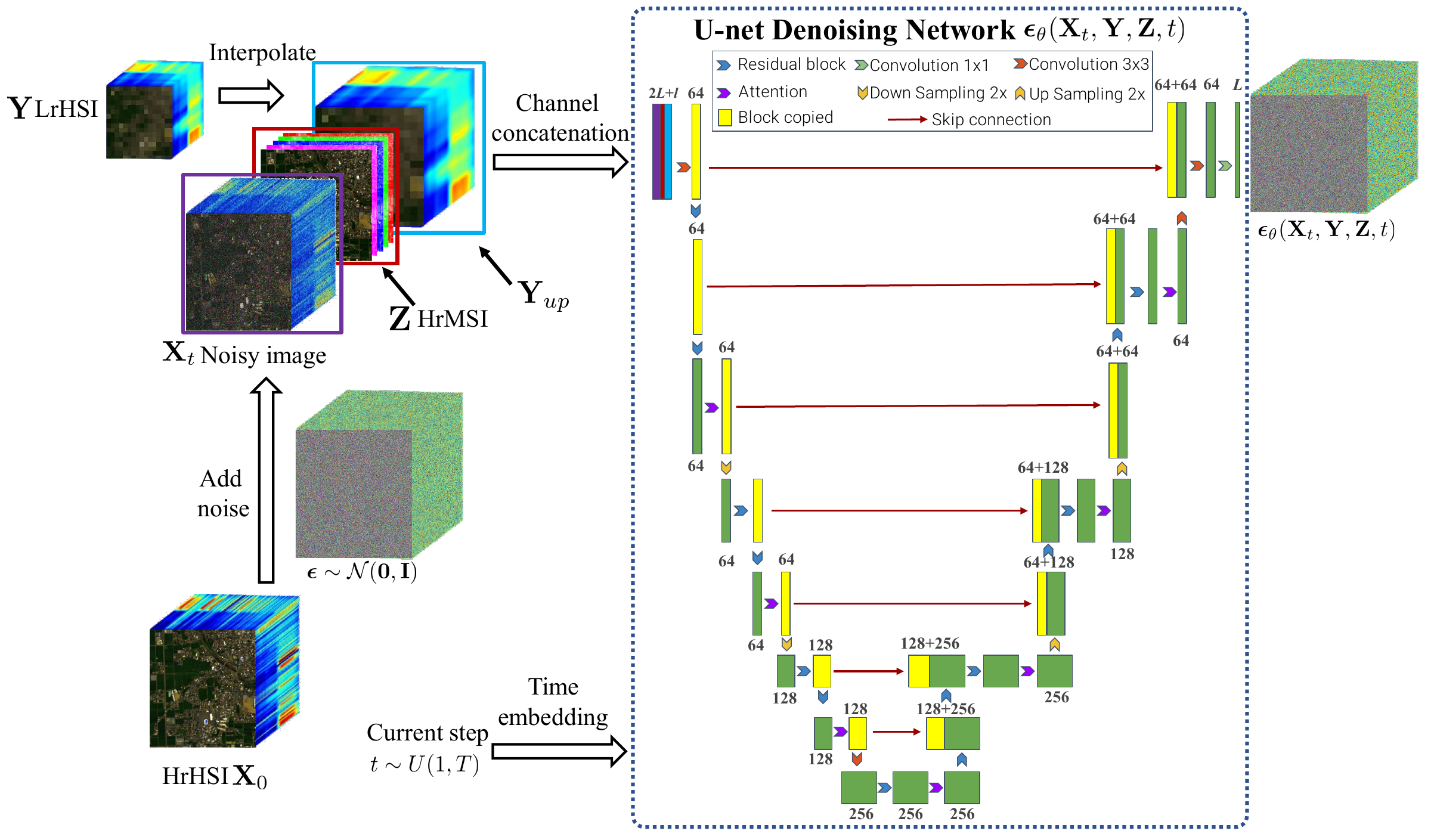}
	\caption{
		Structure of the denoising network $\boldsymbol{\epsilon}_\theta(\mathbf{X}_t,\mathbf{Y},\mathbf{Z},t)$.
		In each training time step $t$, we first obtain $\mathbf{X}_t$ by  adding Gaussian noise to HrHSI $\mathbf{X}_0$
		and interpolate LrHSI $\mathbf{Y}$ to the same spatial resolution of HrMSI $\mathbf{Z}$.
		Then, $\mathbf{X}_t,\mathbf{Z}$ and $\mathbf{Y}_{up}$ will be used as network inputs after the channel dimensions are concatenated.
		The current step $t$ will be input to each residual block in the network in the form of a time embedding.
		In the U-net denoising network, there are a series of convolutional residual blocks, skip connections and attention modules.
		Last, the network predicts the noise that we add before.	
}
	\label{fig.ddpm1}
\end{figure*}

\section{The Proposed DDPM-Fus Model}
In this section, we present the proposed DDPM-Fus, including the problem formulation, forward and backward processes, objective function, optimization and fast fusion strategy.
\label{sec.sec3}

\subsection{Problem Formulation}
The goal of HSI-MSI fusion is to obtain an HrHSI, $\mathcal{X}\in \mathbb{R}^{L\times  W\times H }$, benefiting simultaneously from spatial details of the HrMSI, $\mathcal{Z}\in \mathbb{R}^{ l\times W\times H}$, and spectral signatures of the LrHSI, $\mathcal{Y}\in \mathbb{R}^{L\times  w\times h }$, where  $\{L,W,H,\}$ and $\{l,w,h\}$  denote the channels, widths and heights of high-resolution and low-resolution image cubes, respectively.
It is generally accepted that, $w\ll W, h\ll H$ and $l\ll L$, thus it is nontrivial to perform the image fusion with strong spatial and spectral resolution differences.
The following linear observation model is adopted by most of the fusion literature\cite{cstf,NSSR}:
\begin{align}
	\bf Y&= \mathbf{XBS}+\mathbf{N}_y,\nonumber\\
	\bf Z&= \mathbf{ RX+N}_z,	
	\label{eq.observationModel}
\end{align}
where $\mathbf{X}  \in \mathbb{R}^{L \times WH}$, $\mathbf{Y}  \in \mathbb{R}^{L \times wh}$ and $\mathbf{Z}  \in \mathbb{R}^{l \times WH}$ are 2D matrices unfolding the corresponding 3D tensors of HrHSI, LrHSI and HrMSI, respectively.
$\mathbf{B} \in \mathbb{R}^{WH\times WH} $  is the blurring matrix constructed by the PSF kernel.
$\mathbf{S}\in \mathbb{R}^{WH\times wh}$ represents the spatial downsampling operator which is used combined with the blurring matrix to formulate the spatial degradation process.
$\mathbf{R}  \in \mathbb{R}^{l\times L}$, the SRF matrix, denotes the spectral merging process.
$\mathbf{N} _y$ and $\mathbf{N} _z$ are two Gaussian noise matrices, which are independent of the image data.

The objective is recovering $\mathbf{X}$ with observations $\mathbf{Y},\mathbf{Z}$.
In the first class of fusion methods, conventional unsupervised model, $\mathbf{B,S}$ and $\mathbf{R}$ are often assumed as priors.
While in the second class of fusion methods, supervised deep learning-based models, several HrHSI are known as the ground truths for the training of fusion networks.
Once the training is complete, these models perform the fusion process in other new observation pairs, $\mathbf{Y}$ and $\mathbf{Z}$. 
Our proposed DDPM-Fus is belonging to the second class methods.
We see the HSI-MSI fusion problem as modeling the conditional probabilistic distribution $p(\mathbf{X}|\mathbf{Y},\mathbf{Z})$.
Then one can obtain the desired HrHSI by sampling from this distribution.
Inspired by the advanced denoising diffusion generative model \cite{ddpm}, we build two processes below to learn the above conditional distribution.

\subsection{Forward Diffusion Process}
Following DDPM \cite{ddpm}, we use a Markovian chain of length $T$ with the Gaussian diffusion kernel adding noise to the HrHSI $\mathbf{X}$.
\begin{equation}
	q(\mathbf{X}_t|\mathbf{X}_{t-1})=\mathcal{N}(\sqrt{1-\beta_{t}}\mathbf{X}_{t-1},\beta _t\mathbf{I}),
	\label{eq.diffusionKernel}
\end{equation}
where $t\in \{1,2,\dots,T\}$, $\mathbf{X}_0=\mathbf{X}$ is the ground truth image,
$\mathbf{X}_T$ is the noisy image at when the end of the forward process, $\beta_t \in \{\beta_1,\beta_2,\dots,\beta_T\}$ is a sequence of hyperparameters representing the variance of Guassian noise and $\mathbf I$ is the identity matrix.


Due to the property of Gaussian distribution,
one can directly obtain the output of each diffusion step by
\begin{align}
	q(\mathbf{X}_t|\mathbf{X}_{0})&=\mathcal{N}(\sqrt{\bar\alpha_t}\mathbf{X}_{0},\sqrt{1-\bar\alpha_t}\mathbf{I}) 	\label{eq.forwardGaussian}\\ 
	\text{or}\; \mathbf{X}_t&=\sqrt{\bar {\alpha}_t}\mathbf{X}_0 + \sqrt{1-\bar {\alpha}_t} \boldsymbol{\epsilon},
	\label{eq.forwardGaussianC}
\end{align}
where $\bar {\alpha}_t=\prod_{s=1}^{t}(1-\beta_s) $ and $\boldsymbol{\epsilon}\sim \mathcal{N}(\mathbf{0},\mathbf{I})$.
Note that when $T \rightarrow \infty$, 
$q(\mathbf{X}_ \infty)=\mathcal{N}(\mathbf{0},\mathbf{I})$.

The forward diffusion is shown at the top of Fig.~\ref{fig.ddpm}.

\subsection{Reverse Denoising Process}
After the forward diffusion process, we use the reverse denoising process to recover the HrHSI from the noise input and the pairs of LrHSI and HrMSI.  
Specifically, the reverse process maps from the standard Gaussian noise to the HrHSI $\mathbf{X}$ and we assume the inverse of the above forward diffusion kernel \eqref{eq.diffusionKernel} is another Gaussian as 
\begin{equation}
	p_\theta(\mathbf{X}_{t-1}|\mathbf{X}_t)=\mathcal{N}(\mu_\theta(\mathbf{X}_t,\mathbf{Y},\mathbf{Z},t),\sigma^2_t\mathbf{I}),
	\label{eq.inverseKernel}
\end{equation}
where $\mu_\theta(\cdot)$ is constructed by a deep neural network parameterized by $\theta$ and
 $\sigma^2_t \in \{\sigma^2_1,\sigma^2_2,\dots,\sigma^2_T\}$ is a sequence of hyperparameters, 
which will be given in the Sec.~\ref{sec.subsec3objFunc}.

\subsection{Objective Function}
\label{sec.subsec3objFunc}
In this subsection, we present the objective function of the proposed DDPM-Fus. 
The evidence lower bound (ELBO) of the log-likelihood is 
\begin{equation}
	\mathcal{L}(\theta)=-\sum_{t=1}^T \text{KL}\left(q(\mathbf{X}_{t-1}|\mathbf{X}_{t},\mathbf{X}_{0})||p_\theta(\mathbf{X}_{t-1}|\mathbf{X}_{t})\right),
	\label{eq.elboKL}
\end{equation}
where $\text{KL}(\cdot||\cdot)$ denotes the Kullback-Leibler divergence between two distributions.
The detailed derivation  of \eqref{eq.elboKL} will be given in the APPENDIX.
$q(\mathbf{X}_{t-1}|\mathbf{X}_{t},\mathbf{X}_{0})$ is the posterior distribution which can be deduced to

\begin{align}
q(\mathbf{X}_{t-1}|\mathbf{X}_{t},\mathbf{X}_{0})&\propto q(\mathbf{X}_{t}|\mathbf{X}_{t-1},\mathbf{X}_{0})q(\mathbf{X}_{t-1}|\mathbf{X}_{0})\nonumber\\
&=q(\mathbf{X}_{t}|\mathbf{X}_{t-1})q(\mathbf{X}_{t-1}|\mathbf{X}_{0}).
\label{eq.conditionalPosterior}
\end{align}
Since the conditional distribution \eqref{eq.diffusionKernel} and the prior distribution \eqref{eq.forwardGaussian} here are both Gaussian,
the posterior distribution is also Gaussian due to the self-conjugate nature of the Gaussian distribution \cite{prml}.
Thus the above distribution \eqref{eq.conditionalPosterior} is
\begin{equation}
q(\mathbf{X}_{t-1}|\mathbf{X}_{t},\mathbf{X}_{0}) =\mathcal N (\tilde\mu_t(\mathbf{X}_t,\mathbf{X}_0),
\tilde \beta_t \mathbf{I}),
\label{eq.reverse}
\end{equation}
where 
\begin{align}
	\tilde\mu_t(\mathbf{X}_t,\mathbf{X}_0)&= \frac{\sqrt{1-\beta_t}(1-\bar \alpha _{t-1})}{1-\bar \alpha _t}\mathbf{X}_t+\frac{\sqrt {\bar \alpha_{t-1}}\beta_t}{{1-\bar \alpha _t}} \mathbf{X}_0,\label{eq.predMean}\\
	\tilde \beta_t&=\frac{1-\bar\alpha_{t-1}}{1-\bar\alpha_{t}}\beta_t.
\end{align}

The KL divergence in the ELBO \eqref{eq.elboKL} is tractable because all distributions disturbed here are independent Gaussian.  
Thus the KL divergence at one-time step can be written by
\begin{align}
	\mathcal {L}_t(\theta)
	&=-\text{KL}(q(\mathbf{X}_{t-1}|\mathbf{X}_{t},\mathbf{X}_{0})||p_\theta(\mathbf{X}_{t-1}|\mathbf{X}_{t},\mathbf{Y},\mathbf{Z}))\nonumber\\
	&=-\frac{1}{2\sigma^2_t}\|\tilde\mu_t (\mathbf{X}_t,\mathbf{X}_0)-\mu_\theta(\mathbf{X}_t,\mathbf{Y},\mathbf{Z},t)\|^2_F+C,
	\label{eq.klOneStep}
\end{align}
where $C$ is a constant independent of $\theta$ including some untrainable parameters.
Following \cite{ddpm}, we can use the relationship \eqref{eq.forwardGaussianC} eliminating $\mathbf{X}_0$ in \eqref{eq.predMean} as 
\begin{equation}
	\tilde\mu_t(\mathbf{X}_t,\mathbf{X}_0)= \frac{1}{\sqrt{1-\beta _t}}\left (\mathbf{X}_t-\frac{\beta_t}{\sqrt{1-\bar\alpha _t}}\boldsymbol{\epsilon}\right ).
\end{equation}

Meanwhile, the prediction of the neural network $\mu_\theta(\mathbf{X}_t,\mathbf{Y},\mathbf{Z},t)$ can be writen as the same form 
\begin{align}
	&\mu_\theta(\mathbf{X}_t,\mathbf{Y},\mathbf{Z},t)\nonumber\\
= 	&\frac{1}{\sqrt{1-\beta _t}}\left(\mathbf{X}_t-\frac{\beta_t}{\sqrt{1-\bar\alpha _t}}\boldsymbol{\epsilon}_\theta(\mathbf{X}_t,\mathbf{Y},\mathbf{Z},t)\right).
	\label{eq.reverseFusion}
\end{align}
where $\boldsymbol{\epsilon}_\theta(\cdot)$ is another form of $\mu_\theta(\cdot)$, which predicts the noise added in the forward process and will be introduced in the next subsection.

Thus the KL divergence \eqref{eq.klOneStep} can be further simplified to
\begin{equation}
	\mathcal {L}_t(\theta)=- \frac{\beta_t^2}{2\sigma^2_t (1-\beta_t)(1-\bar\alpha_t)}
	\|\boldsymbol{\epsilon}-\boldsymbol{\epsilon}_\theta(\mathbf{X}_t,\mathbf{Y},\mathbf{Z},t)\|^2_F+C.
\end{equation}
In the training phase, we set $\sigma^2_t=\tilde \beta_t$, so $C=0$.
Last, we can optimize the ELBO \eqref{eq.elboKL} step-by-step along time and in each step we focus on the simple loss function ignoring hyperparameter coefficients as 
\begin{equation}
	\mathcal {L}_{\text{simple}}(\theta)=  
	\|\boldsymbol{\epsilon}-\boldsymbol{\epsilon}_\theta(\mathbf{X}_t,\mathbf{Y},\mathbf{Z},t)\|^p_p,
	\label{eq.lossSimple}
\end{equation}
where typically $p=1$ or $2$ for using $\ell_1$ and $\ell_2$ loss.
In our experiments, we use $\ell_1$ loss and the ablation study about $p$ will be given in the Sec.~\ref{sec.modelDisscussion}.

\begin{algorithm}[!t]
	\small
	\SetAlgoLined
	\caption{Training algorithm of DDPM-Fus}
	\label{al.DDPM}
	\LinesNumbered  
	\KwIn{paired training data: ($\mathbf{Y}$, $\mathbf{Z}$, $\mathbf{X}_0$)\;
		hyperparameter sequence: $\{\beta_1,\beta_2,\dots,\beta_T\},$ 
	}
	Initialize $\theta$ randomly\;
	\Repeat{\textbf{training phase end}}{
	 $\boldsymbol{\epsilon}\sim \mathcal{N}(\mathbf{0},\mathbf{I})$\;
	  $t\sim U(1,T)$\;
	  Compute the noisy image by \eqref{eq.forwardGaussianC}: $\mathbf{X}_t=\sqrt{\bar {\alpha}_t}\mathbf{X}_0 + \sqrt{1-\bar {\alpha}_t} \boldsymbol{\epsilon}$\\
		Compute the gradient of \eqref{eq.lossSimple} w.r.t. $\theta$:
		$\nabla_\theta\|\boldsymbol{\epsilon}-\boldsymbol{\epsilon}_\theta(\mathbf{X}_t,\mathbf{Y},\mathbf{Z},t)\|$\;
		Update ${\theta}$ via Adam optimizer \cite{adam}.\\
	}

\end{algorithm}

\begin{algorithm}[!t]
	\small
	\SetAlgoLined
	\caption{Fusion of DDPM-Fus }
	\label{al.DDPMtest}
	\LinesNumbered  
	\KwIn{paired test LrHSI and HrMSI: ($\mathbf{Y}^\prime$, $\mathbf{Z}^\prime$)\;
		the sub-sequence of $\{1,2,\dots,T\}$ is denoted as $\{\tau _1,\tau_2,\dots,\tau_d =T\}$
	}
 
	$\mathbf{X}_T^\prime\sim \mathcal N(\mathbf{0},\mathbf{I})$\;
	\For{$t=\tau_d ,\dots,\tau _1$}	 {$\boldsymbol{\epsilon}\sim \mathcal{N}(\mathbf{0},\mathbf{I})$ if $t>1$, else $\boldsymbol{\epsilon}=\mathbf{0}$\\
		Compute the estimated noise by $\boldsymbol{\epsilon}_\theta(\mathbf{X}_t^\prime,\mathbf{Y}^\prime,\mathbf{Z}^\prime,t)$\;
		Sample the denoising output $\mathbf{X}_{t-1}^\prime$ from   DDIM \cite{ddim} sampler \eqref{eq.ddim}, \eqref{eq.ddim2} and \eqref{eq.ddim3}\;
  }
\end{algorithm}

\subsection{Network Optimization}
Next, we construct the conditional denoising network $\boldsymbol{\epsilon}_\theta(\cdot)$ using the U-net architecture. Overall, the network is shown in Fig.~\ref{fig.ddpm1}.
Inputs of the U-net consists of 4 parts, $\{\mathbf{X}_t,\mathbf{Y},\mathbf{Z},t\}$.
We have already compute $\mathbf{X}_t$ by \eqref{eq.forwardGaussianC} in the forward diffuison process. 
Then we interpolate LrHSI $\mathbf{Y}$ using bicubic method to the same spatial resolution as HrMSI $\mathbf{Z}$  and concatenate it with HrMSI and noisy image along the channel dimension.
These data preprocess can be formulated by 
\begin{align}
	\mathbf{Y}_{up}&=\text{Bicubic} (\mathbf{Y}) \uparrow_{S\times},\\
	\mathbf{IN}&=[\mathbf{X}_t,\mathbf{Z},\mathbf{Y}_{up}].
\end{align}
For simplicity, assuming that the spatial resolution of $\mathbf{X}$ is divisible by $\mathbf{Y}$, $S=W/w=H/h$.
$\uparrow_{S\times}$ represents increasing the spatial resolution of LrHSI $S$ times by bicubic interpolation.
 $[\cdot]$ denotes the concatenation operator.
$\mathbf{IN}$ is the data input of the following denoising network.
A U-net is used here to implement the denoising process,
which consists of several convolutional residual blocks and attention layers.
For clarity, the channel numbers of some main feature maps have been marked on the corresponding layers and the forward computation process is illustrated in Fig.~\ref{fig.ddpm1}.
Besides, the current time step $t$ is considered  as another condition and is input to each residual block in the form of learned embedding vectors, such as the positional embedding in the Transformer \cite{transformer}.
To specify, the training process of the proposed DDPM-Fus is summarized in Algorithm \ref{al.DDPM}.

\subsection{DDPM-Fus with Fast Sampling}
Once training is complete, we can obtain fusion results on the test data by implementing the reverse denoising process.
Instead of sampling from \eqref{eq.inverseKernel} step-by-step, we get the fused results via 
another reverse sampler to reduce the fusion time. 
This sampler was proposed in \cite{ddim},  where a non-Markovian forward process was designed and allowed to skip steps in the implementation of the reverse process, namely denoising diffusion implicit models (DDIM).
Specifically,
the new forward process can be formulated by
\begin{align}
&q_\sigma (\mathbf{X}_{t-1}|\mathbf{X}_{t},\mathbf{X}_{0})\nonumber\\
=&\mathcal{N}(
\sqrt{\bar{\alpha}_{t-1}}\mathbf{X}_{0}+ \sqrt{1-\bar\alpha_{t-1}-\sigma_t^2}  \cdot  \frac{\mathbf{X}_t-\sqrt{\bar{\alpha}}_t\mathbf{X}_{0}}{\sqrt{1-\bar\alpha_t}},\sigma^2_t \mathbf{I}).
\label{eq.ddim}
\end{align}
Based on the above relation, we can use the estimated values of $\mathbf{X}_0$ and noises to implement the reverse process as 
\begin{align}
\mathbf{X}_0 &\approx \frac{\mathbf{X}_t-\sqrt{1-\bar\alpha_t}      \boldsymbol{\epsilon}_\theta(\mathbf{X}_t,\mathbf{Y},\mathbf{Z},t)}{\sqrt{\bar{\alpha}}_t},\label{eq.ddim2}\\
\frac{\mathbf{X}_t-\sqrt{\bar{\alpha}}_t\mathbf{X}_{0}}{\sqrt{1-\bar\alpha_t}}&\approx\boldsymbol{\epsilon}_\theta(\mathbf{X}_t,\mathbf{Y},\mathbf{Z},t). 
\label{eq.ddim3}
\end{align}

To specify, we can obtain the fused results on the test data by implementing the reverse denoising process concluded in Algorithm \ref{al.DDPMtest}.

\begin{figure}[!t]
	\centering
	
	\hspace{-0.4cm}
	\subfigure[]{
		\includegraphics[height=2 cm]{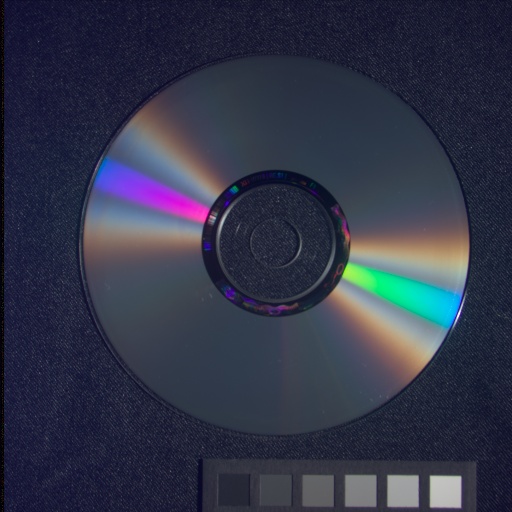}
	}\hspace{-0.2cm}
	\subfigure[]{
		\includegraphics[height=2cm]{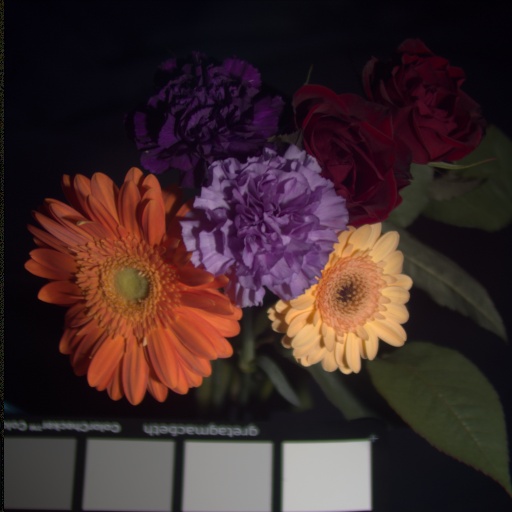}
	}\hspace{-0.2cm}
	\subfigure[]{
		\includegraphics[height=2cm]{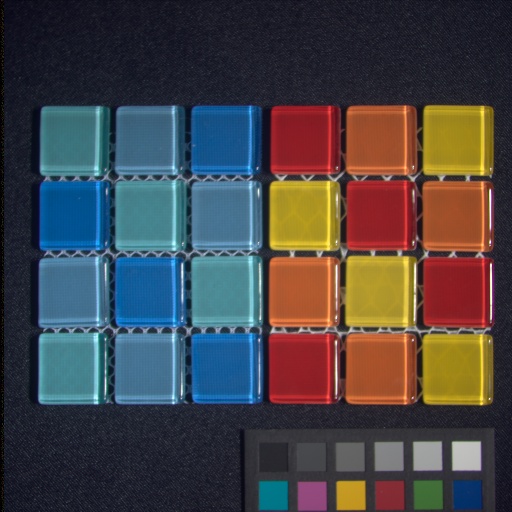}
	}\hspace{-0.2cm}
	\subfigure[]{
		\includegraphics[height=2cm]{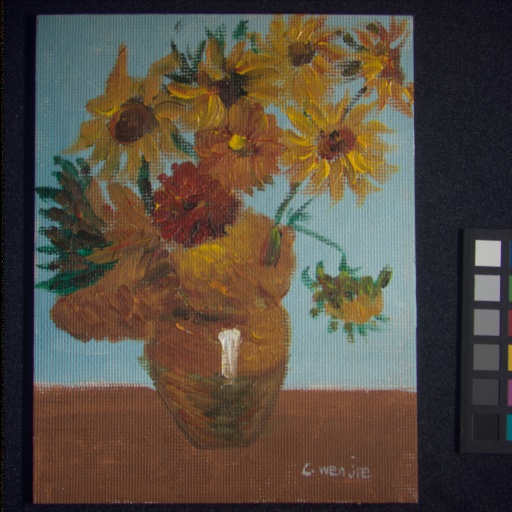}
	}
	
	\hspace{-0.5cm}
	\subfigure[]{
		\includegraphics[height=2.2cm]{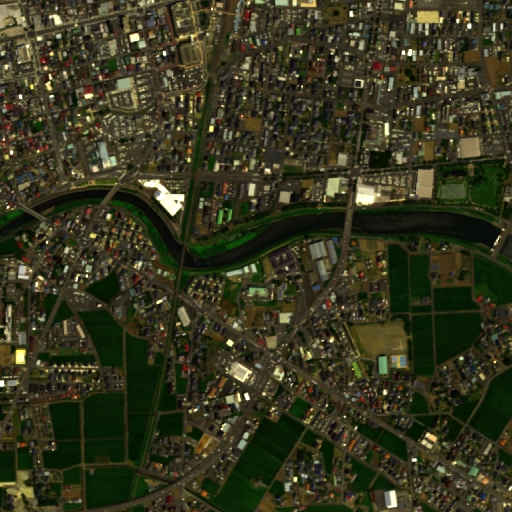}
	}\hspace{-0.2cm}
	\subfigure[]{
		\includegraphics[height=2.2cm]{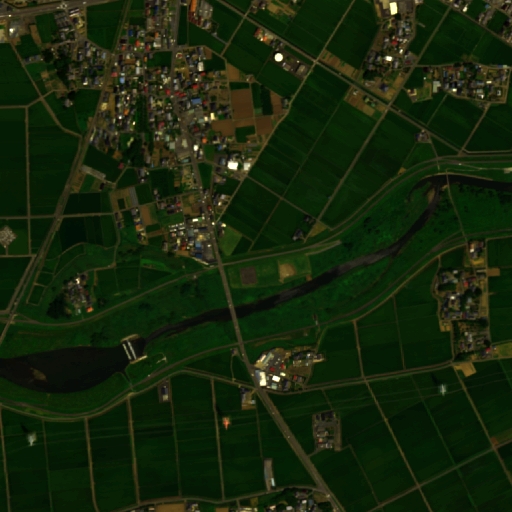}
	}\hspace{-0.2cm}
	\subfigure[]{
		\includegraphics[height=2.2cm]{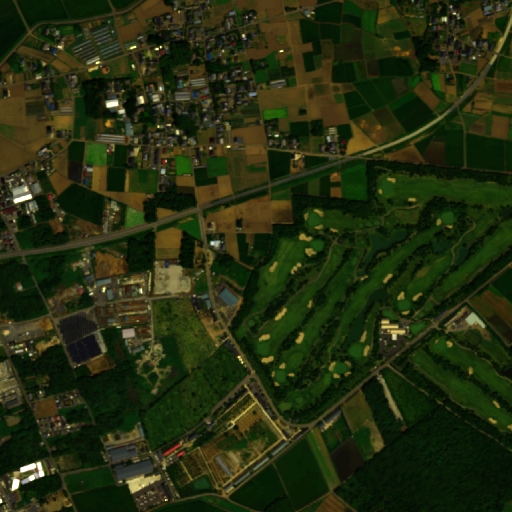}
	}\hspace{-0.2cm}
	\subfigure[]{
		\includegraphics[height=2.2cm]{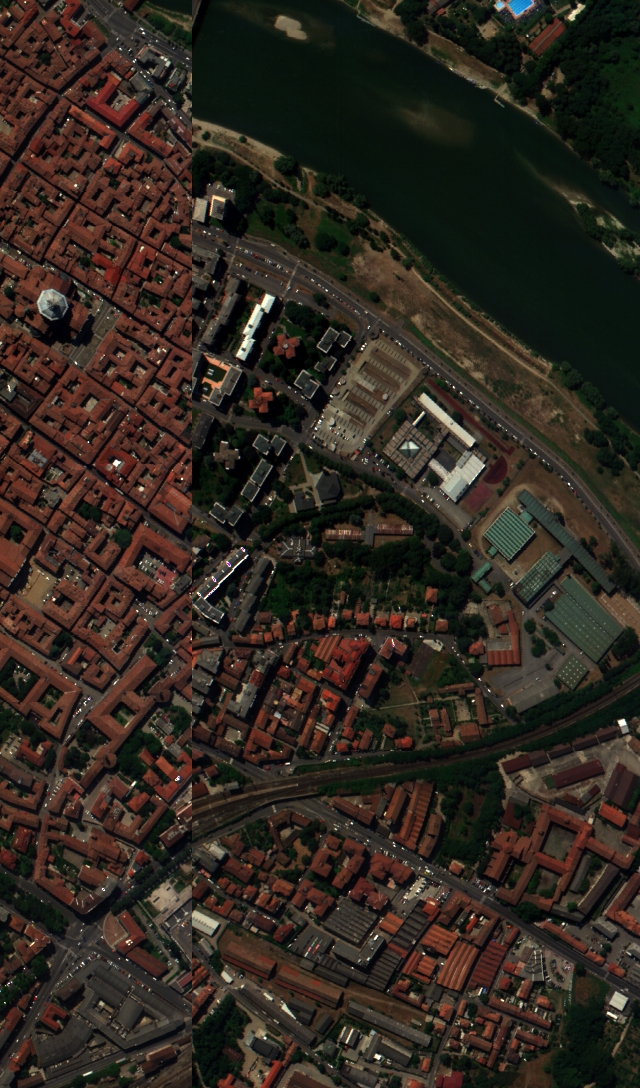}
	}
	\caption{Some benchmark RGB images in datasets, \textbf{CAVE}: (a) \textit{CD}, (b) \textit{flowers}, (c) \textit{glass tiles} and (d) \textit{oil painting}, \textbf{Chikusei}: (e) \textit{region2}, (f) \textit{region6} and (g) \textit{region11}, \textbf{Pavia Center}: (h) \textit{Pavia Center}.
	}
	\label{fig.benchmarkimages}
\end{figure}

\begin{table}[!t]
	\centering
	\renewcommand\arraystretch{1.2}
	\caption{Quantitative metrics of the comparison methods on the 10 test images from the CAVE dataset. The best results are in bold, while the second best methods are underlined.}
	\label{tab.CAVE}
	\begin{tabular}{c|c|c|c|c}
		\toprule[1.3pt]
		Methods  & PSNR  & SAM & ERGAS  & SSIM      \\\hline\hline
		PANnet &   30.68& 13.68   & 1.11    & 0.864    \\
		HSRnet & 41.05 & \underline{7.94}  & \underline{0.41} & \underline{0.975}    \\\hline
     	SSRnet & \underline{41.73} & 8.17  & 0.41 &\underline{0.975}                   \\
		MHFnet & 36.46 & 22.12 & 1.64 & 0.951\\
		EDBIN & 40.68 & 8.96  & 0.48 & 0.969      \\
    	DDPM-Fus & \textbf{43.66} & \textbf{5.69}  & \textbf{0.34} &\textbf{0.986}      \\\hline\hline
		Ideal value & +$\infty$&0&0&1\\
		\bottomrule[1.3pt]           
	\end{tabular}
\end{table}

\section{Experiments}
\label{sec.sec4}
We demonstrate the experimental results of our proposed DDPM-Fus conducted on the CAVE \cite{cave}\footnote{https://www.cs.columbia.edu/CAVE/databases/multispectral/} dataset and two  remote sensing datasets, namely, Chikusei \cite{chikusei}\footnote{http://naotoyokoya.com/Download.html} and Pavia Center\footnote{https://rslab.ut.ac.ir/data}.
For a fair comparison, several algorithms of supervised deep learning-based HSI-MSI fusion methods are implemented with their open-source code, including
PANnet\cite{pannet} \footnote{https://xueyangfu.github.io/projects/iccv2017.html}, 
HSRnet\cite{HSRnet}\footnote{https://github.com/liangjiandeng/HSRnet}, 
SSRnet\cite{ssrnet}\footnote{https://github.com/hw2hwei/SSRNET},
MHFnet\cite{mhfnet}\footnote{https://github.com/XieQi2015/MHF-net}
and 
EDBIN\cite{EDBIN} \footnote{https://github.com/wwhappylife/Deep-Blind-Hyperspectral-Image-Fusion}.


\begin{figure*}[!tb]
	\hspace{-0.6cm}
	\begin{tabular}{c}	
		\includegraphics[height=2.9cm,trim= 60 0 60 0,clip]{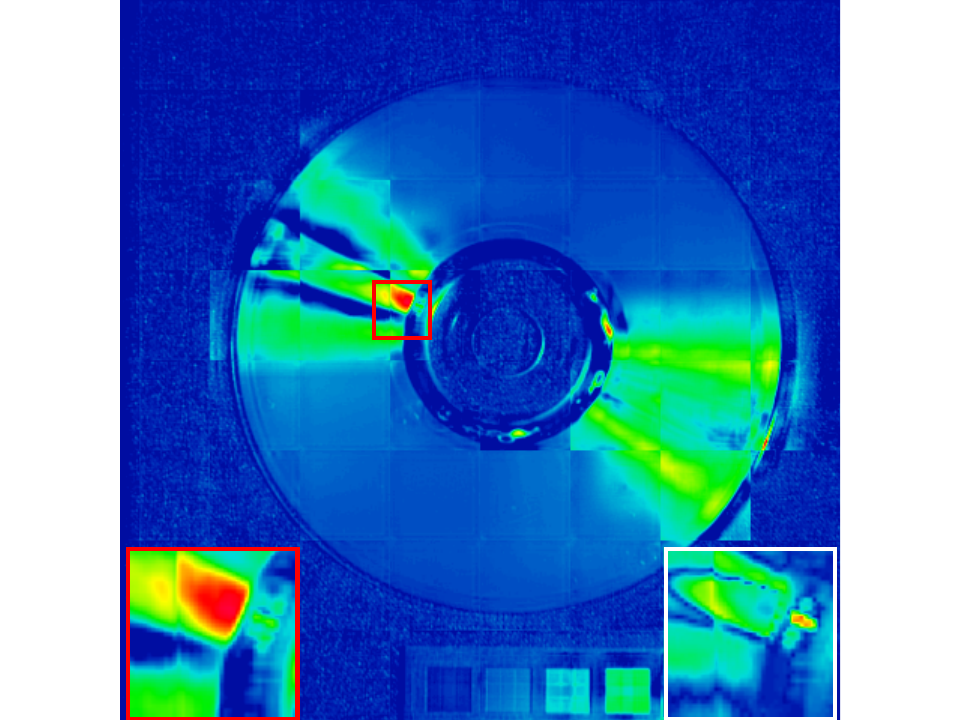}
		\\(a) $\text{PANnet}$
		\\(\textit{29.56/9.51})  
	\end{tabular}\hspace{-0.5cm}
	\begin{tabular}{c}	
		\includegraphics[height=2.9cm,trim= 60 0 60 0,clip]{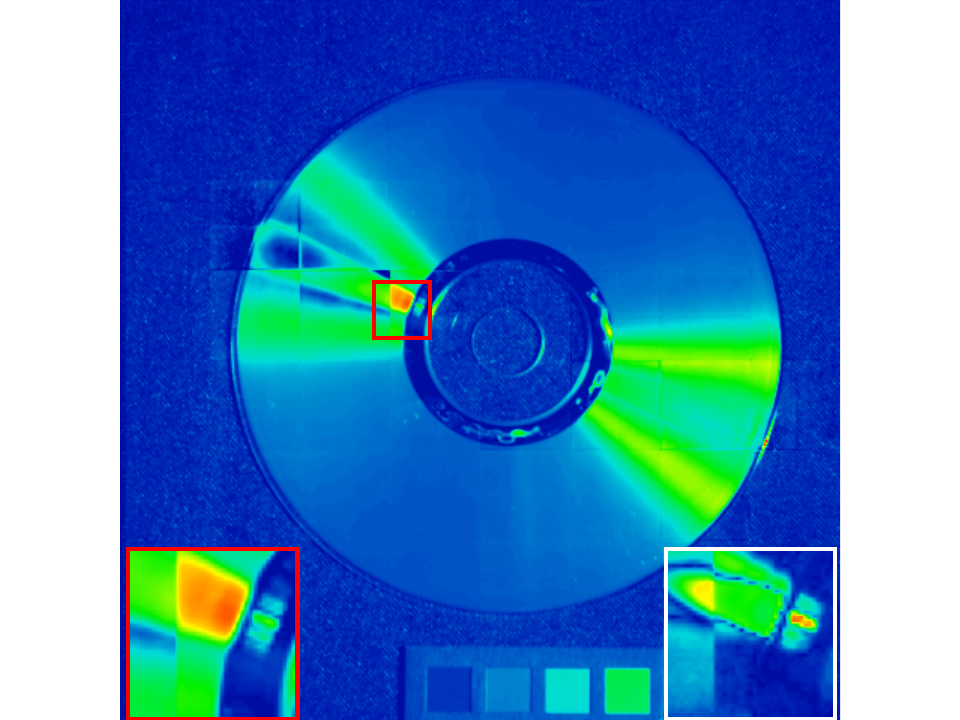}
		\\(b) $\text{HSRnet}$
		\\(\textit{\underline{32.44/5.66}})  
	\end{tabular}\hspace{-0.5cm}
	\begin{tabular}{c}	
		\includegraphics[height=2.9cm,trim= 60 0 60 0,clip]{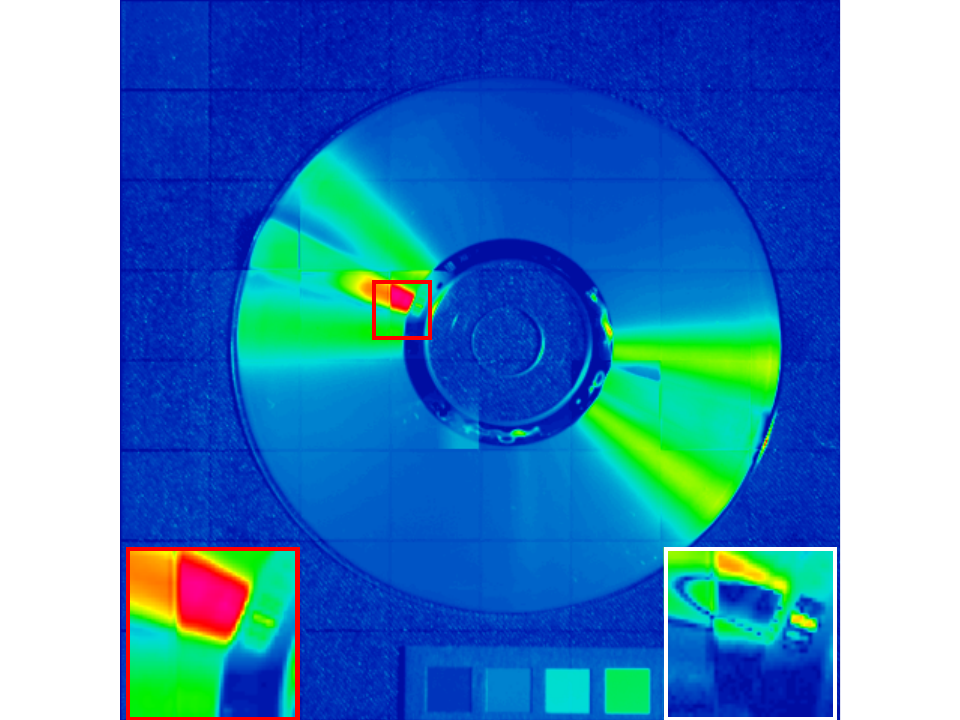}
		\\(c) $\text{SSRnet}$
		\\(\textit{31.97/7.04})  
	\end{tabular}\hspace{-0.5cm}
	\begin{tabular}{c}	
		\includegraphics[height=2.9cm,trim= 60 0 60 0,clip]{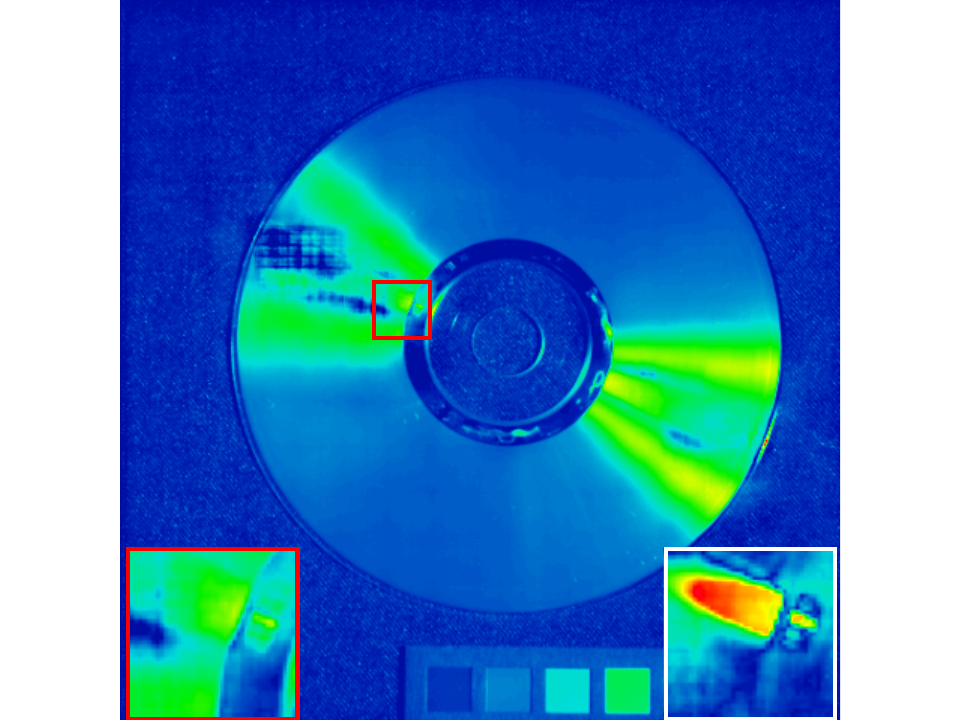}
		\\(d) $\text{MHFnet}$
		\\(\textit{30.16/17.30})  
	\end{tabular}\hspace{-0.5cm}

	\begin{tabular}{c}	
		\includegraphics[height=2.9cm,trim= 60 0 60 0,clip]{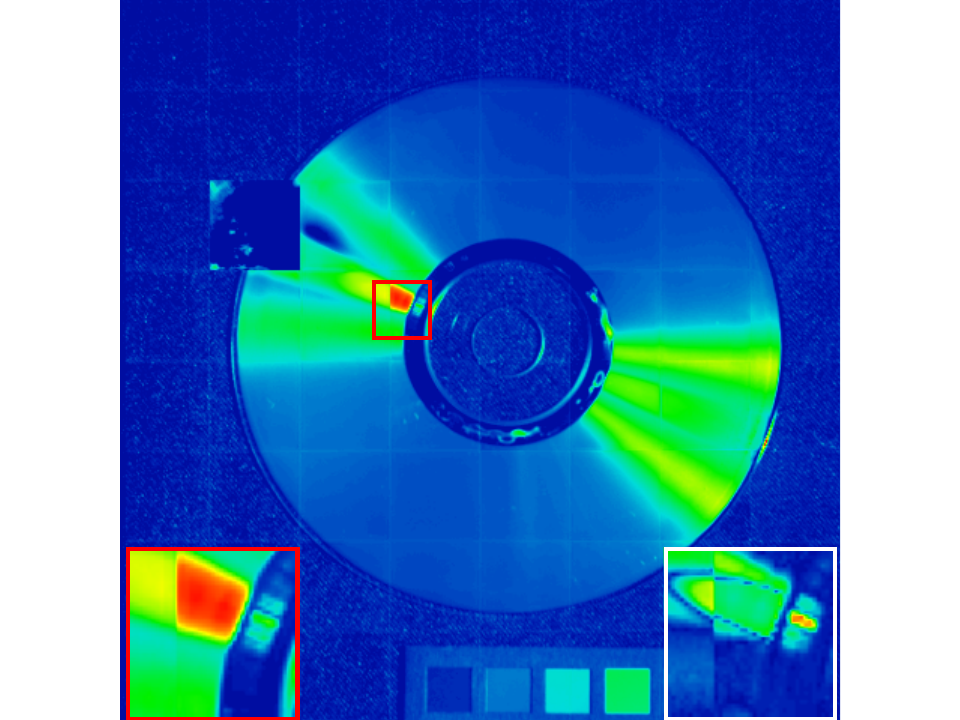}
		\\(e) $\text{EDBIN}$
		\\(\textit{29.11/7.32})  
	\end{tabular}\hspace{-0.5cm}
	\begin{tabular}{c}	
		\includegraphics[height=2.9cm,trim= 60 0 60 0,clip]{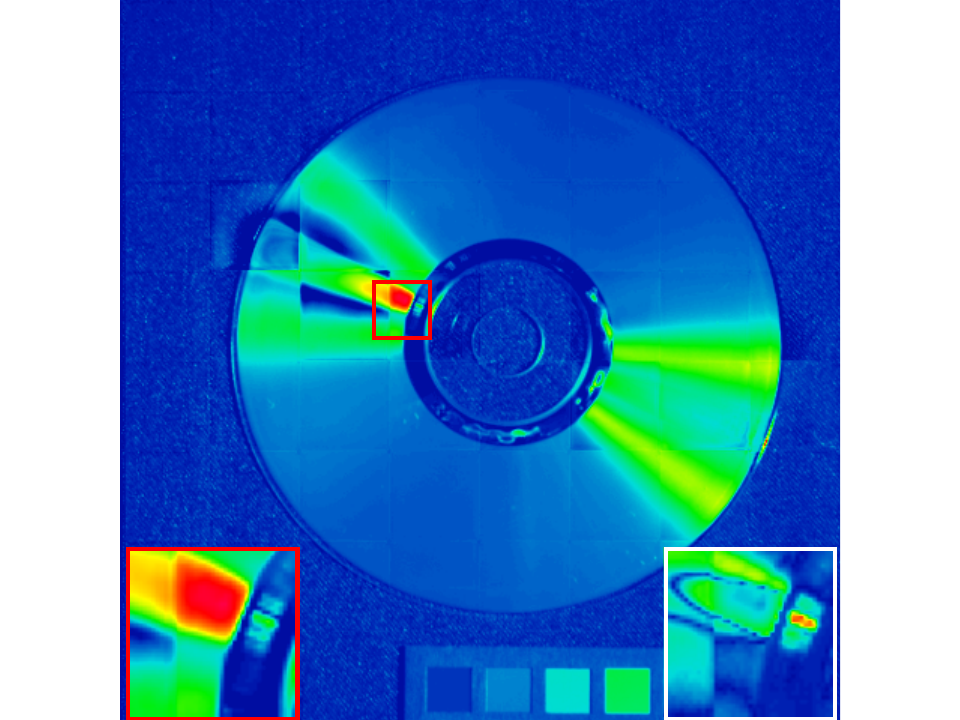}
		\\(f) $\text{DDPM-Fus}$
		\\(\textit{\textbf{32.45/4.95}})  
	\end{tabular}
	\vspace{0.2cm}
	\hspace{-0.6cm}
	\begin{tabular}{c}	
		\includegraphics[height=2.9cm,trim= 60 0 60 0,clip]{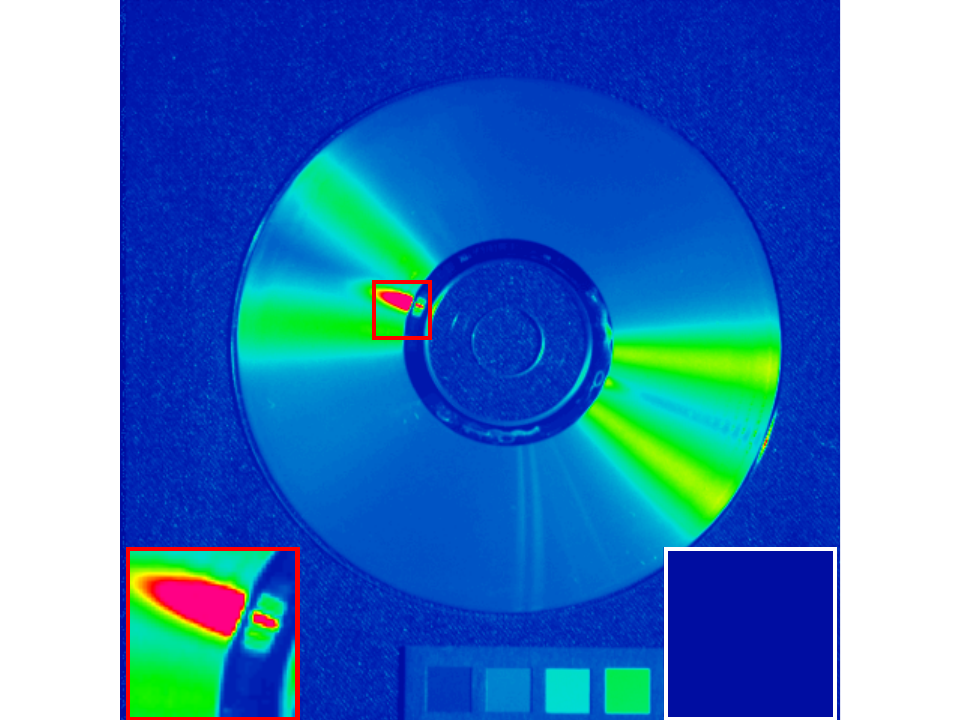}
		\\(g) $\text{GT}$
		\\(\textit{PSNR/SAM})  
	\end{tabular}\hspace{-0.5cm}
	\begin{tabular}{c}	
		\includegraphics[height=2.9cm,width=1.5cm]{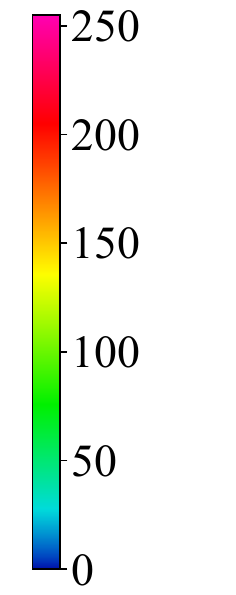}
	\end{tabular}
	\subfigure{
		\begin{tabular}{c}
			\includegraphics[height=5cm,trim=10 0 10 0,clip]{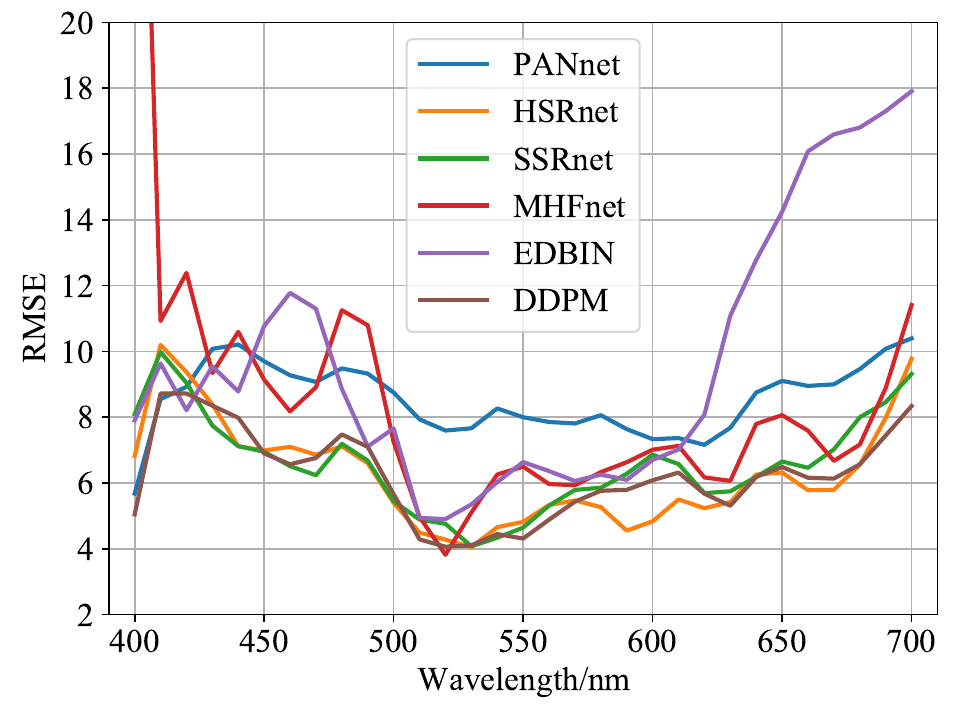}\\
			(h) 
		\end{tabular}
	}
	
	\caption{(a-g) The 21st band (600 nm) of fused HrHSI (\textit{CD} in the CAVE dataset) obtained by the testing methods, where a ROI zoomed in 9 times (bottom-left) and the corresponding residual maps (bottom-right) are shown for detail visualization. PSNR and SAM are also listed for comparison. (h) The corresponding RMSE along with spectral bands.
	}
	\label{fig.CAVE.cd}
\end{figure*}

\begin{figure*}[!tb]
	\hspace{-0.6cm}
	\begin{tabular}{c}	
		\includegraphics[height=2.9cm,trim= 60 0 60 0,clip]{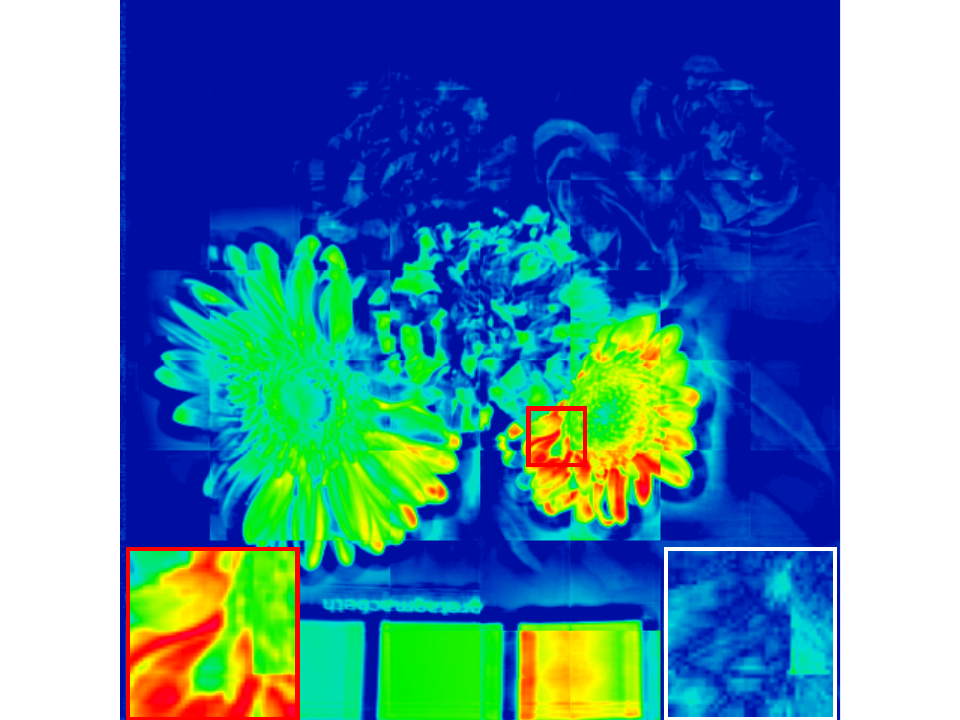}
		\\(a) $\text{PANnet}$
		\\(\textit{31.82/18.07})  
	\end{tabular}\hspace{-0.5cm}
	\begin{tabular}{c}	
		\includegraphics[height=2.9cm,trim= 60 0 60 0,clip]{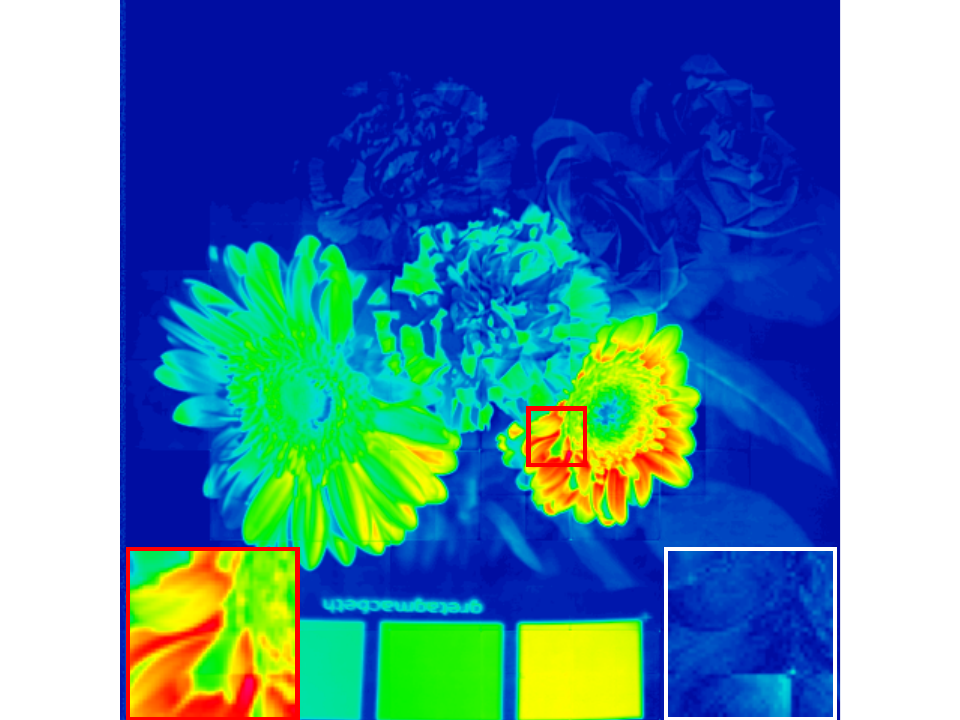}
		\\(b) $\text{HSRnet}$
		\\(\textit{42.73/12.22})  
	\end{tabular}\hspace{-0.5cm}
	\begin{tabular}{c}	
		\includegraphics[height=2.9cm,trim= 60 0 60 0,clip]{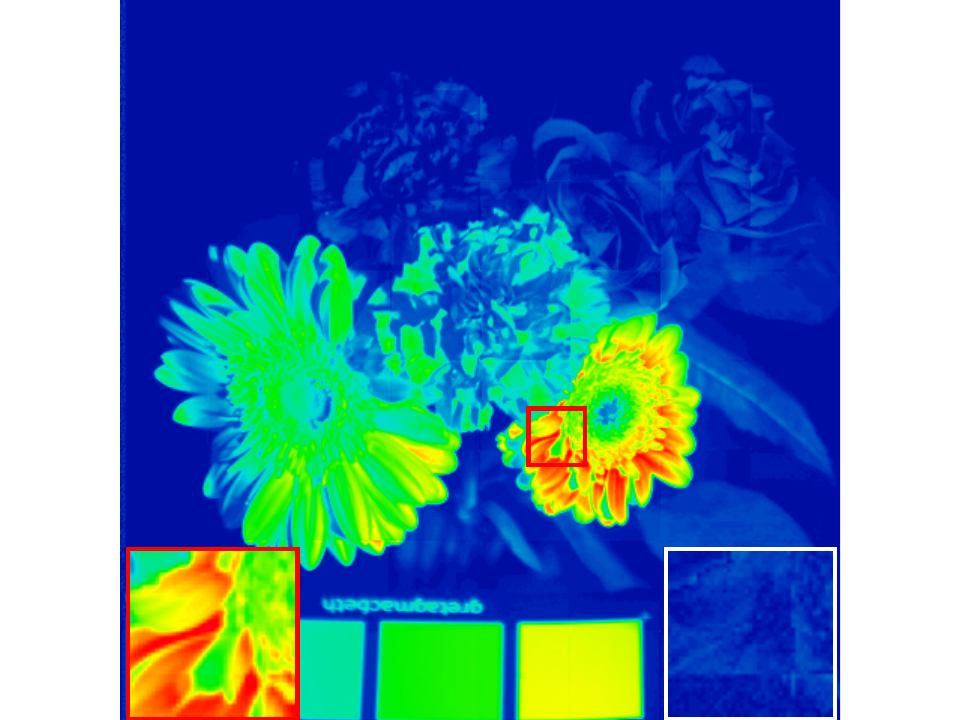}
		\\(c) $\text{SSRnet}$
		\\(\textit{\underline{42.76/11.11}})  
	\end{tabular}\hspace{-0.5cm}
	\begin{tabular}{c}	
		\includegraphics[height=2.9cm,trim= 60 0 60 0,clip]{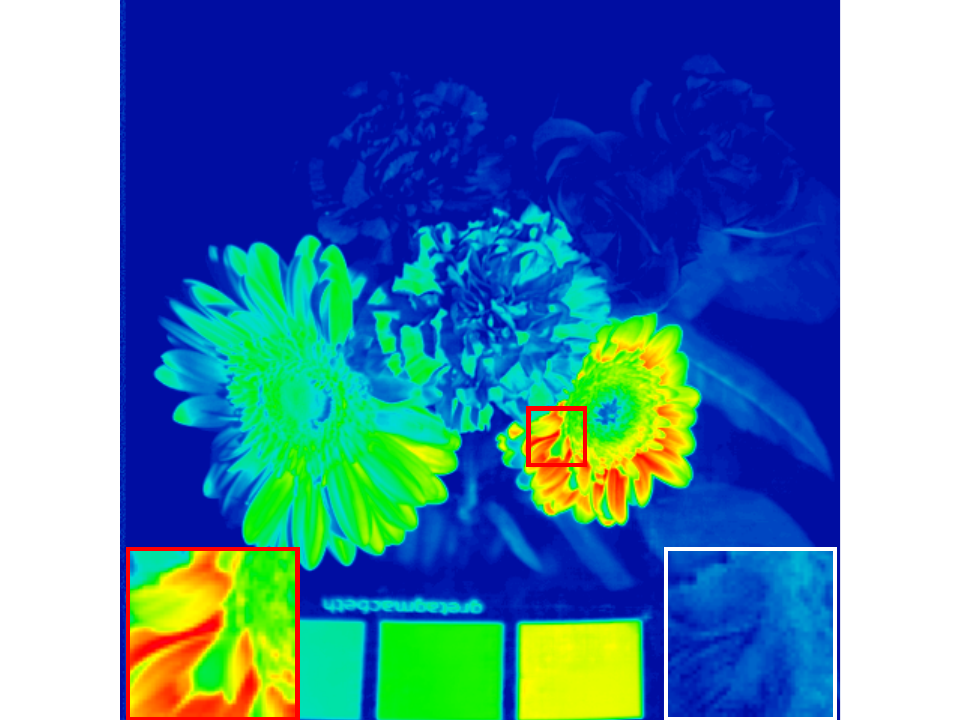}
		\\(d) $\text{MHFnet}$
		\\(\textit{32.16/23.75})  
	\end{tabular}\hspace{-0.5cm}
	
	\begin{tabular}{c}	
		\includegraphics[height=2.9cm,trim= 60 0 60 0,clip]{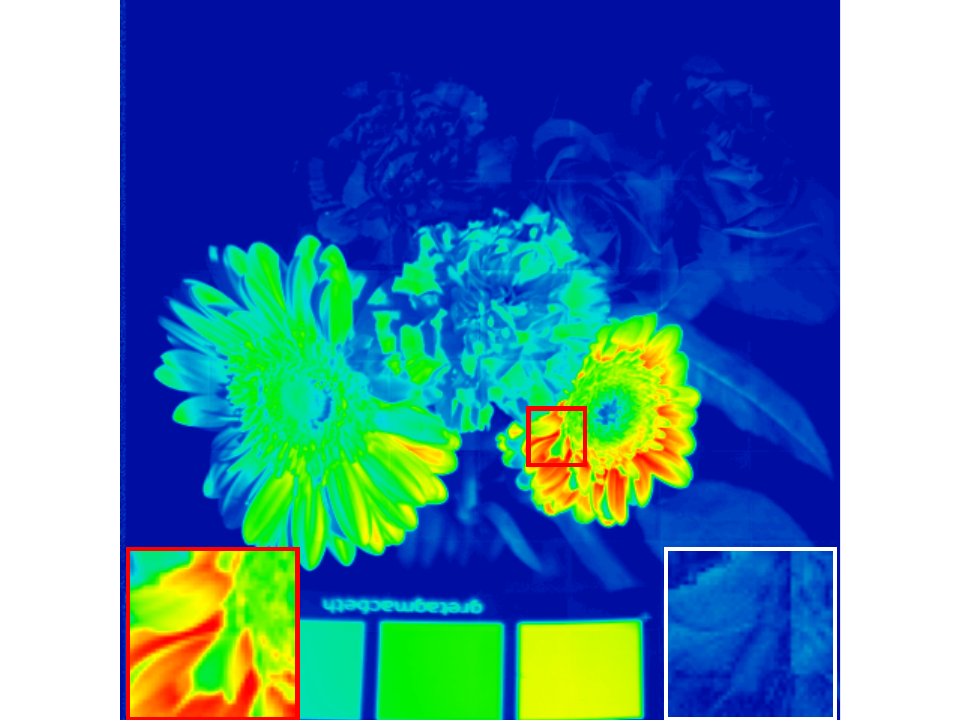}
		\\(e) $\text{EDBIN}$
		\\(\textit{42.09/11.97})  
	\end{tabular}\hspace{-0.5cm}
	\begin{tabular}{c}	
		\includegraphics[height=2.9cm,trim= 60 0 60 0,clip]{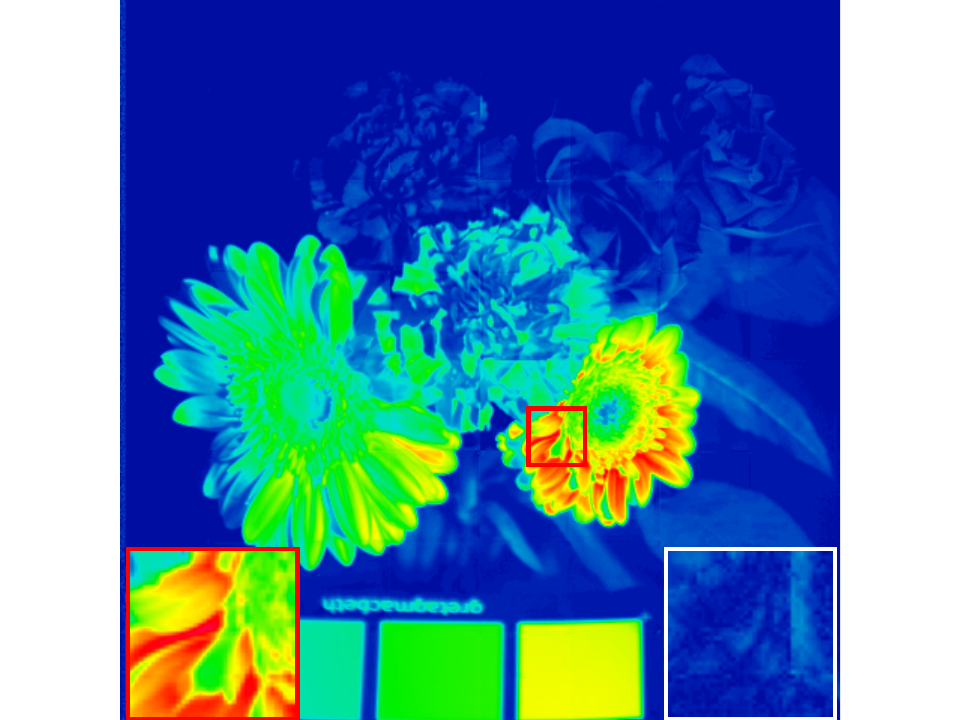}
		\\(f) $\text{DDPM-Fus}$
		\\(\textit{\textbf{44.20/8.48}})  
	\end{tabular}
	\vspace{0.2cm}
	\hspace{-0.6cm}
	\begin{tabular}{c}	
		\includegraphics[height=2.9cm,trim= 60 0 60 0,clip]{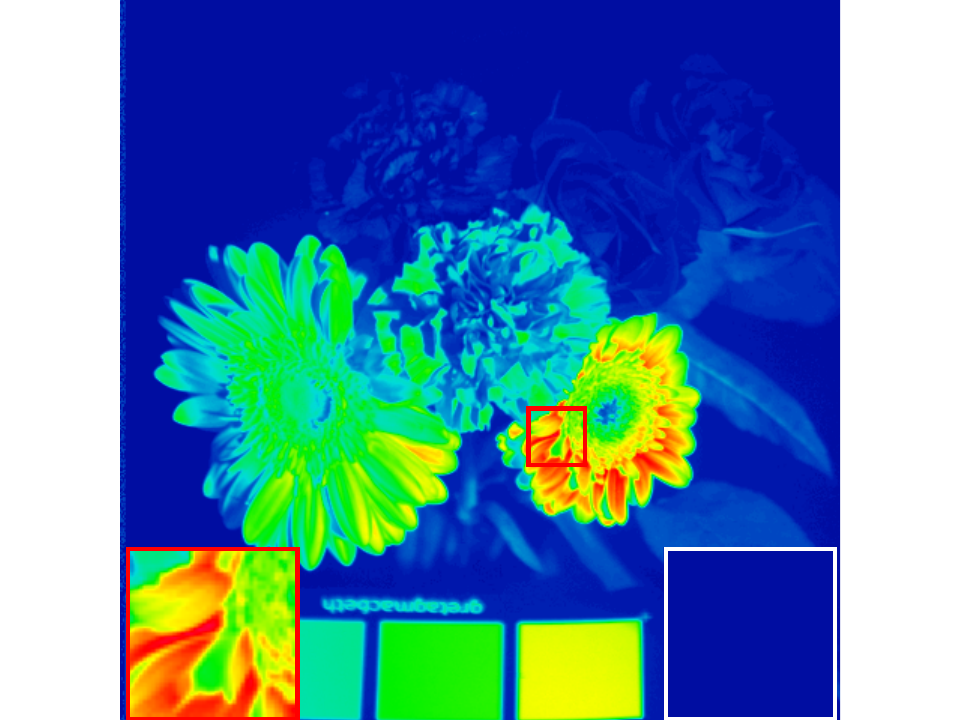}
		\\(g) $\text{GT}$
		\\(\textit{PSNR/SAM})  
	\end{tabular}\hspace{-0.5cm}
	\begin{tabular}{c}	
		\includegraphics[height=2.9cm,width=1.5cm]{fig/cycFusion/visualres/colorbar1}
	\end{tabular}
		\subfigure{
		\begin{tabular}{c}
			\includegraphics[height=5cm,trim= 10 0 10 0,clip]{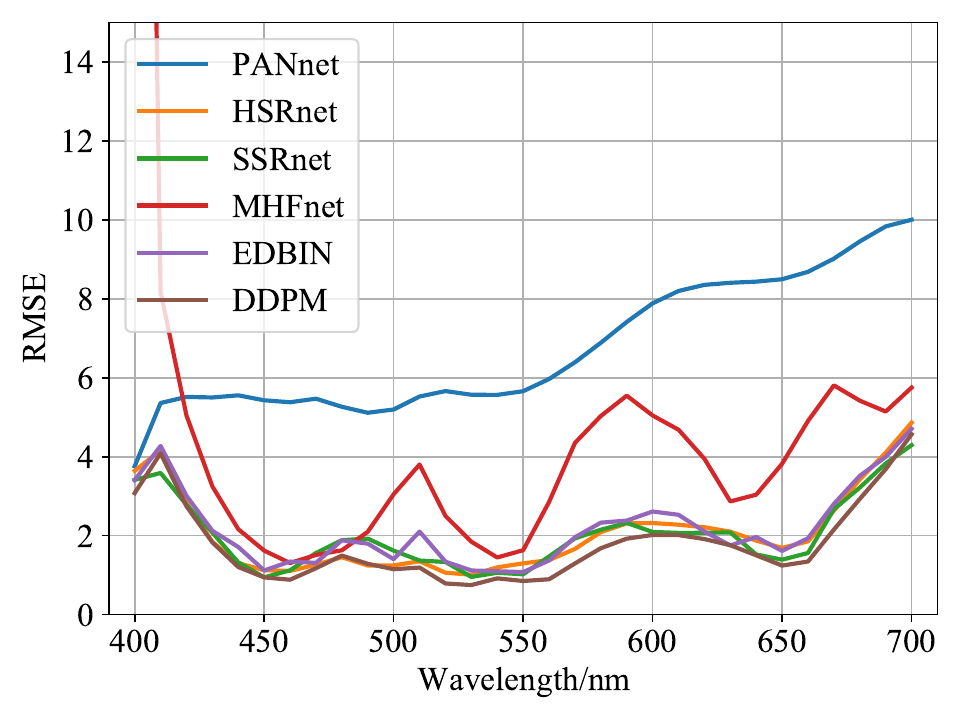}\\
			(h) 
		\end{tabular}
	}
	
	\caption{(a-g) The 21st band (600 nm) of fused HrHSI (\textit{flowers} in the CAVE dataset) obtained by the testing methods, where a ROI zoomed in 9 times (bottom-left) and the corresponding residual maps (bottom-right) are shown for detail visualization. PSNR and SAM are also listed for comparison. (h) The corresponding RMSE along with spectral bands.
	}
	\label{fig.CAVE.flowers}
\end{figure*}


\subsection{Data Description}
The first  dataset is CAVE collected by Apogee Alta U260 camera contains  32  scenes of size $512\times 512$ with  31  spectral bands covering a wavelength range of 400 nm to 700 nm at 10 nm steps.
Followed by \cite{mhfnet} and \cite{HSRnet}, we set 20 images as training data and the other 12 images as test data.
The second database is Chikusei acquired by the Headwall Hyperspec-VNIR-C imaging sensor over Chikusei, Japan.
The  original image comprises 128 bands ranging from 363 nm to 1018 nm with a ground sampling distance (GSD) of 2.5 m, and it is $2517\times 2335$ in spatial size.
For convenience, we choose 16 non-overlapped subregions from the raw data for the study, each of which has a 512 by 512 size.
We set 10 images as training data and other 6 images as test data.
The last image we used is acquired by the ROSIS sensor over Pavia, Italy, which contains $ 1096 \times 715$ pixels and 102 bands covering from  430 nm to 860 nm.
A $1088\times 448$ sub-image and another $1088\times 192$ region from it are selected as the training data  and test data, respectively, for the evaluation.
We refer to  the above data as the ground truth (GT) of HrHSI to compare the fusion performance.
Several benchmark images in these  datasets are illustrated in Fig.~\ref{fig.benchmarkimages}. 

Following  \cite{HSRnet, nvpgm} and \cite{fusionNet}, we generate the observed LrHSI by directly averaging the $32\times 32$ spatially disjoint blocks in the HrHSI.
For the CAVE, Chikusei and Pavia Center datasets, we use  the SRF of Nikon D700 camera\footnote{https://www.maxmax.com/spectral\_response.htm}, LANDSAT-8\footnote{https://landsat.gsfc.nasa.gov/article/preliminary-spectral-response-of-the-operational-land-imager-in-band-band-average-relative-spectral-response} and an IKONOS-like sensor to simulate the HrMSI via HrHSI, respectively.
The HrMSI of the CAVE dataset has three bands that correspond to the red, green, and blue channels, whereas the HrMSI of the other two remote sensing datasets has four bands, one more NIR band than the former.

\begin{figure*}[!tb]
	\hspace{-0.6cm}
	\begin{tabular}{c}	
		\includegraphics[height=2.9cm,trim= 60 0 60 0,clip]{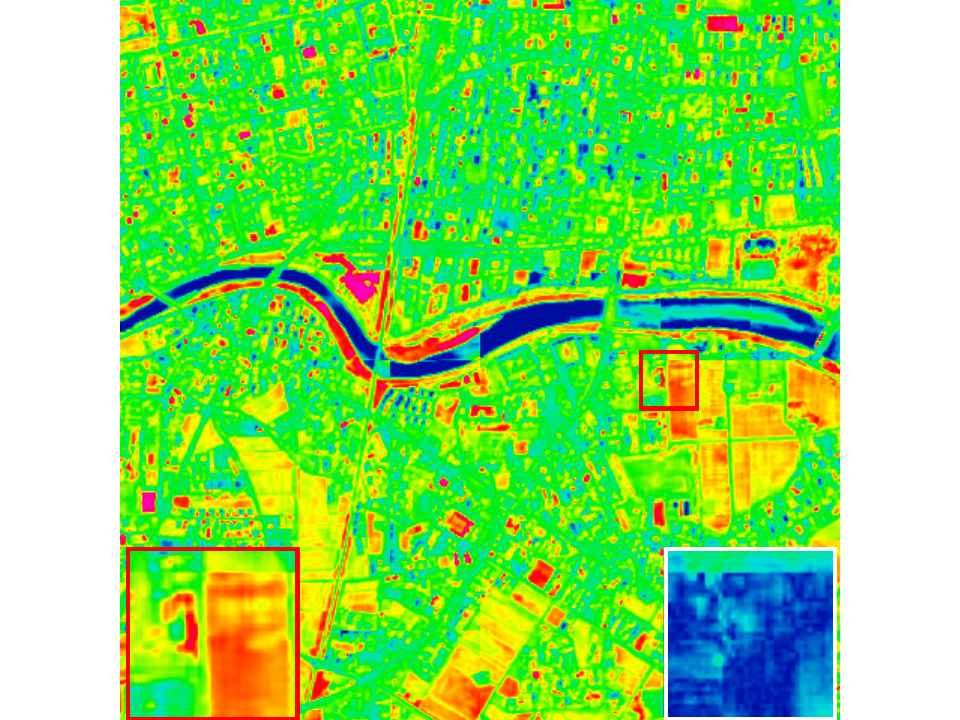}
		\\(a) $\text{PANnet}$
		\\(\textit{26.22/7.04})  
	\end{tabular}\hspace{-0.5cm}
	\begin{tabular}{c}	
		\includegraphics[height=2.9cm,trim= 60 0 60 0,clip]{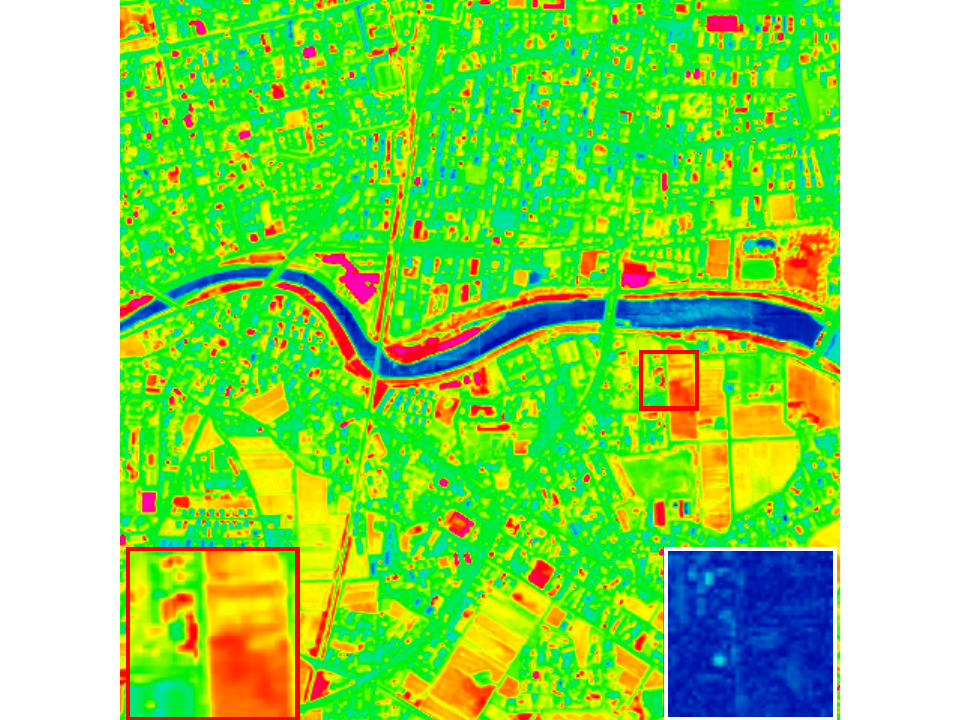}
		\\(b) $\text{HSRnet}$
		\\(\textit{37.67/2.60})  
	\end{tabular}\hspace{-0.5cm}
	\begin{tabular}{c}	
		\includegraphics[height=2.9cm,trim= 60 0 60 0,clip]{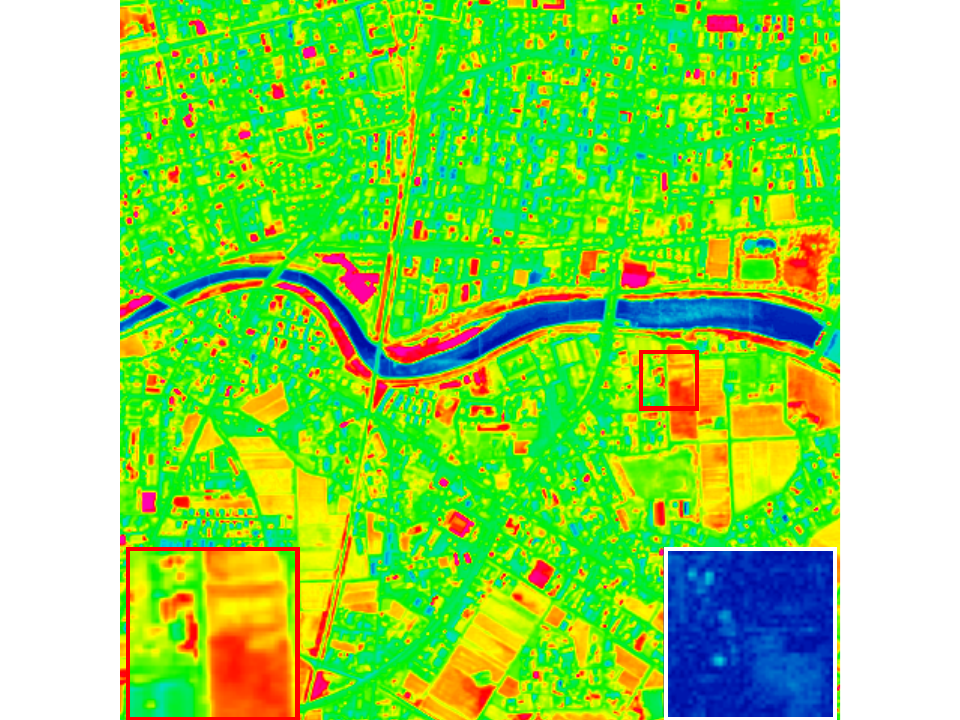}
		\\(c) $\text{SSRnet}$
		\\(\textit{\underline{38.74/2.51}})  
	\end{tabular}\hspace{-0.5cm}
	\begin{tabular}{c}	
		\includegraphics[height=2.9cm,trim= 60 0 60 0,clip]{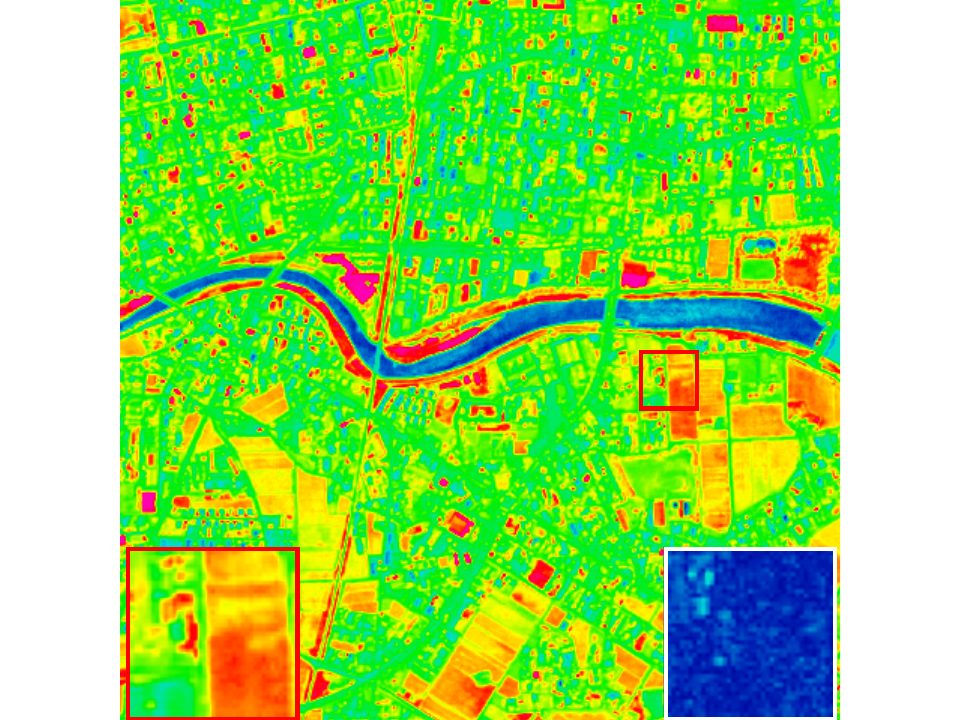}
		\\(d) $\text{MHFnet}$
		\\(\textit{38.09/3.16})  
	\end{tabular}\hspace{-0.5cm}
	
	\begin{tabular}{c}	
		\includegraphics[height=2.9cm,trim= 60 0 60 0,clip]{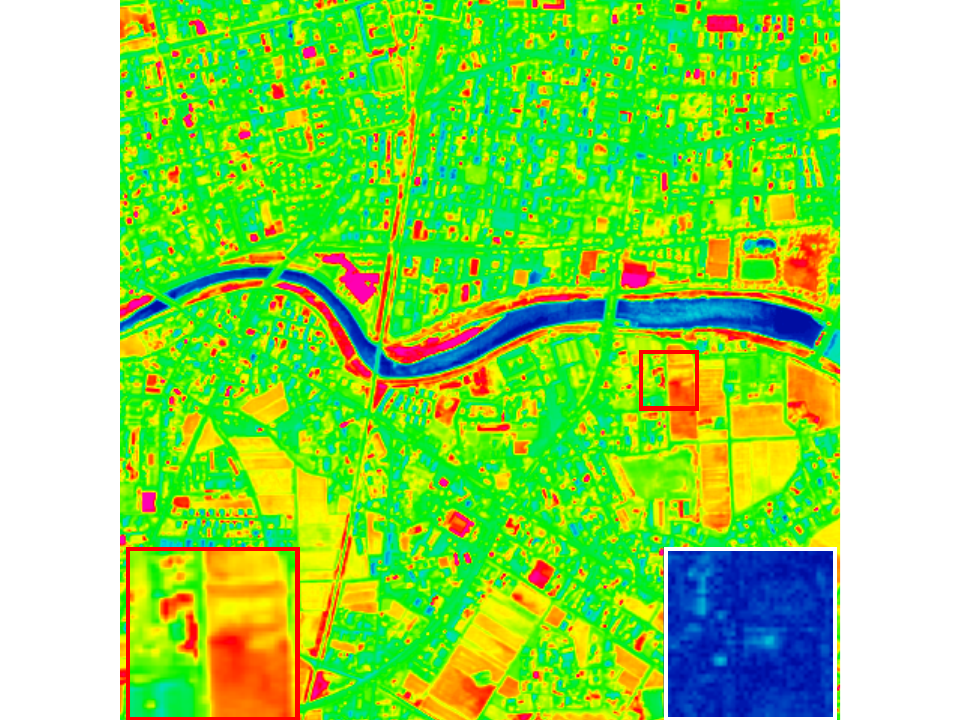}
		\\(e) $\text{EDBIN}$
		\\(\textit{38.73/2.99})  
	\end{tabular}\hspace{-0.5cm}
	\begin{tabular}{c}	
		\includegraphics[height=2.9cm,trim= 60 0 60 0,clip]{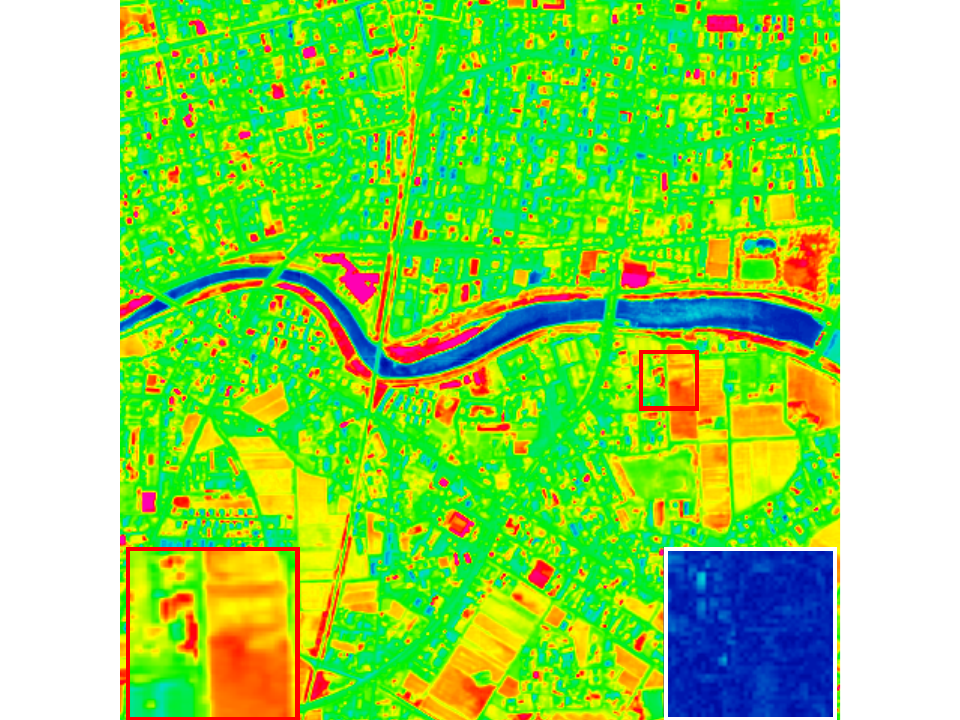}
		\\(f) $\text{DDPM-Fus}$
		\\(\textit{\textbf{39.25/2.42}})  
	\end{tabular}
	\vspace{0.2cm}
	\hspace{-0.6cm}
	\begin{tabular}{c}	
		\includegraphics[height=2.9cm,trim= 60 0 60 0,clip]{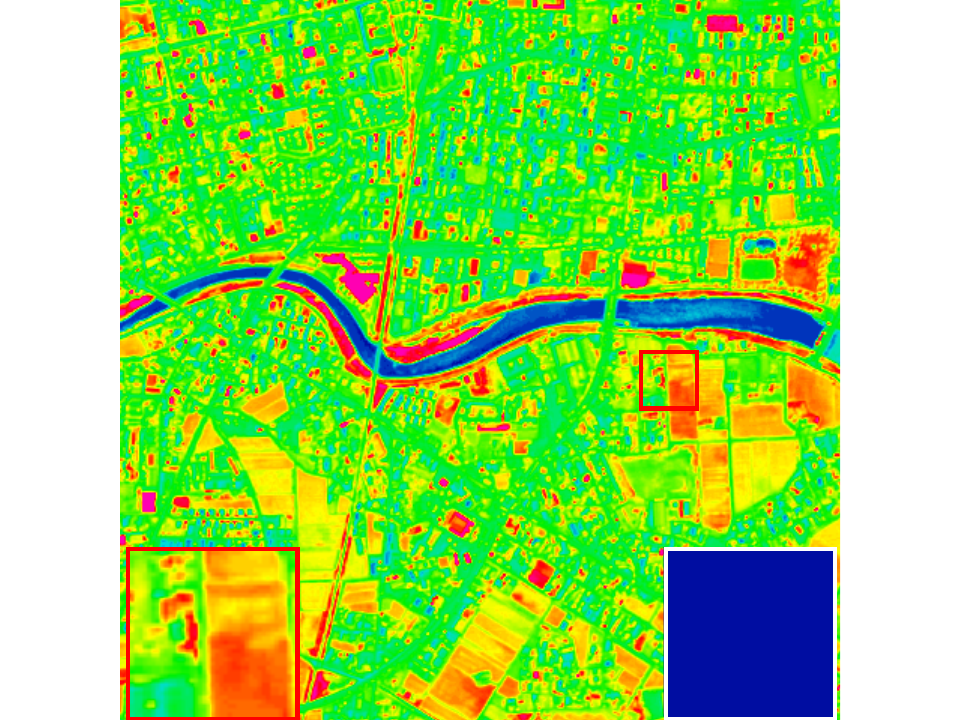}
		\\(g) $\text{GT}$
		\\(\textit{PSNR/SAM})  
	\end{tabular}\hspace{-0.5cm}
	\begin{tabular}{c}	
		\includegraphics[height=2.9cm,width=1.5cm]{fig/cycFusion/visualres/colorbar1}
	\end{tabular}
		\subfigure{
		\begin{tabular}{c}
			\includegraphics[height=5cm,trim= 10 0 10 0,clip]{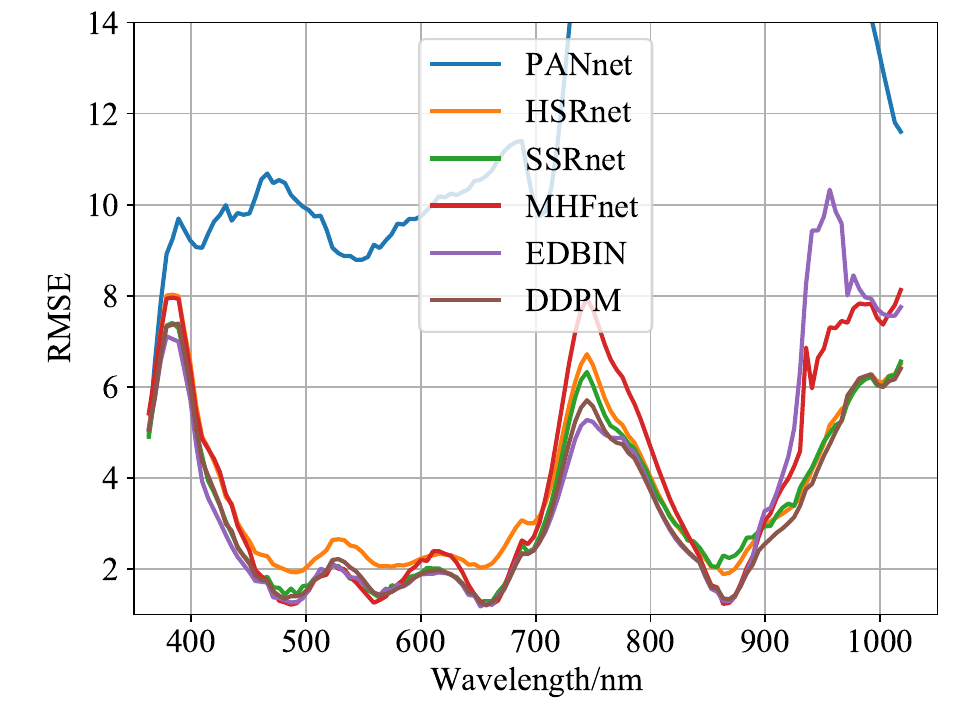}\\
			(h) 
		\end{tabular}
	}
	
	\caption{(a-g) The 73rd band (734 nm) of fused HrHSI (\textit{region2} in the Chikusei dataset) obtained by the testing methods, where a ROI zoomed in 9 times (bottom-left) and the corresponding residual maps (bottom-right) are shown for detail visualization. PSNR and SAM are also listed for comparison.
	(h) The corresponding  RMSE along with spectral bands.
	}
	\label{fig.Chikusei.region2}
\end{figure*}

\begin{table*}[]
	\centering
	\renewcommand\arraystretch{1.2}
	\caption{Quantitative metrics of the comparison methods on the 6 test images of the Chikusei and the test image of the Pavia Center dataset. The best results are in bold, while the second best methods are underlined.}
	\label{tab.Chikusei}
	\begin{tabular}{c|c|c|c|c||c|c|c|c}
		\toprule[1.3pt]
		\multirow{2}{*}{Methods}&\multicolumn{4}{c||}{Chikusei}   & \multicolumn{4}{c}{Pavia Center}\\\cline{2-9}
		& PSNR  & SAM  & ERGAS & SSIM  & PSNR  & SAM  & ERGAS & SSIM \\\hline\hline
		PANnet   & 28.17 & 4.31 & 0.76 & 0.896 & 31.63 & 6.83 & 0.56 & 0.940 \\
		HSRnet   & 38.87 & 2.00 & 0.42 & 0.972 & 44.21 & 3.37 & 0.22 & 0.982 \\\hline
		SSRnet   & \underline{39.66} & \underline{1.89} & \textbf{0.39} & \underline{0.974} & 45.11 & \underline{2.98} & \underline{0.19}& \underline{0.986}  \\
		MHFnet   & 39.04 & 2.33 & 0.44 & 0.963 & 42.82 & 4.37 & 0.26 & 0.977 \\
		EDBIN    & 38.70 & 2.89 & 0.43 & 0.973 & \underline{45.35} & \textbf{2.93} & \textbf{0.18} & \textbf{0.987}\\
		DDPM-Fus & \textbf{40.25} & \textbf{1.86} &\underline{0.41} & \textbf{0.975} & \textbf{45.39} & 3.01 & \underline{0.19} &\underline{0.986} \\\hline\hline
		Ideal value & +$\infty$&0&0&1& +$\infty$&0&0&1\\
		\bottomrule[1.3pt]           
	\end{tabular}
	
\end{table*}

\subsection{Experimental Setup}
\subsubsection{Hyperparameter Settings}
The same architecture of the proposed DDPM-Fus is used for fusion  three datasets with the diffusion time step $T$ is set to $ 2000 $, which is shown in Fig.~\ref{fig.ddpm1}. 
The hyperparameter sequence $\{\beta_1,\beta_2,\dots,\beta_T\}$ is set to a sequence with uniform growth from $ 0 $ to $ 0.01 $. 
The number of training iterations is set to $250k$ and  the Adam optimizer\cite{adam} is adopted here.
In the training phase, we use the cosine annealing learning schedule \cite{sgdr} with the cycle set to $50k$ steps to  promote the convergence, where the maximum learning rate is set to $ 0.0001 $.  
We perform the fusion task on one NVIDIA GeForce RTX 3090 GPU with 24GB memory.
Due to the memory limitation, we divide the training images into small patches with the size of $64\times 64$ and the batch size is set to 8. 

\subsubsection{Performance Metrics}
We use the peak signal-to-noise ratio (PSNR), the relative dimensionless global error in synthesis (ERGAS)\cite{ERGAS}, the spectral angle mapper (SAM), and the structure similarity to quantitatively evaluate the fused results of comparison methods (SSIM) \cite{SSIM}.
PSNR is equivalent to the root mean squared error (RMSE). 
ERGAS is the average relative RMSE of each channel, which can be used to eliminate intensity effects. 
SAM compares the similarity of spectra in radian units. 
The structural similarity between the ground truth and the estimated image is measured using the SSIM criterion, which is widely used in image processing. 
All evaluation criteria are assessed in the 8-bit range, i.e., [0-255].

\subsection{Experiments on the Indoor Dataset}
Table~\ref{tab.CAVE} shows the average performance metrics on the CAVE dataset for all comparison models.
Overall, PANnet and HSRnet as one-stage fusion methods produce poor results.
Compared with them, the multi-stage fusion methods 
show competitive performance, especially SSRnet, bringing 0.68 dB improvement in PSNR compared to HSRnet.
The existing multi-stage fusion models usually contain several stages, such as coarse fusion, spatial correction and spectral recovery in SSRnet, or scores of fusion steps in MHFnet and EDBIN.
Our proposed DDPM-Fus have
thousands of fusion stages and we obtain the approximated fused results by the DDIM sampler \cite{ddim}. 
Unsurprisingly, the proposed DDPM-Fus outperforms all other fusion models
and improves the PSNR by 1.93dB over the second-best model, SSRnet, on the CAVE dataset. 
 
Fig.~\ref{fig.CAVE.cd} and Fig.~\ref{fig.CAVE.flowers} 
 show the fused HrHSI in the 21st band (600 nm) for CD and flowers, respectively. Two regions of interest (ROIs) are
highlighted for comparing detailed differences of all testing methods. 
It can be observed that there are prominent artifacts in the fused image provided by PANnet and EDBIN as shown in 
Fig.~\ref{fig.CAVE.cd} (a) and (e). 
Besides, the progressive fusion methods, SSRnet and DDPM-Fus, obtain lower error maps than other methods as shown in \ref{fig.CAVE.flowers} (c) and (f), respectively.
Clearly, the proposed DDPM-Fus produces the best fusion results of the target HrHSI and corresponding lower absolute error maps than other comparison models.
The band-by-band root mean squared error (RMSE) of all methods on these two images are shown in Fig.~\ref{fig.CAVE.cd} (h)
and 
Fig.~\ref{fig.CAVE.flowers} (h) to compare the reconstruction in each band obtained by the testing algorithms.
These RMSE results further demonstrate the superiority of our proposed model in spectral reconstruction.

\begin{figure}[!tb]

	\begin{tabular}{c}	
		\includegraphics[ width=2cm, trim= 190 0 190 0,clip]{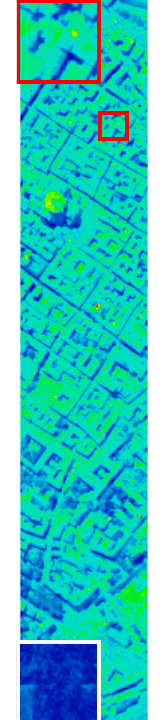}
		\\(a) $\text{PANnet}$
		\\(\textit{31.63/6.83})  
	\end{tabular}\hspace{-0.5cm}
	\begin{tabular}{c}	
		\includegraphics[ width=2cm,trim= 190 0 190 0,clip]{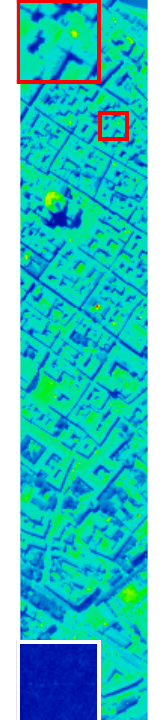}
		\\(b) $\text{HSRnet}$
		\\(\textit{44.21/3.37})  
	\end{tabular}\hspace{-0.5cm}
	\begin{tabular}{c}	
		\includegraphics[ width=2cm,trim= 190 0 190 0,clip]{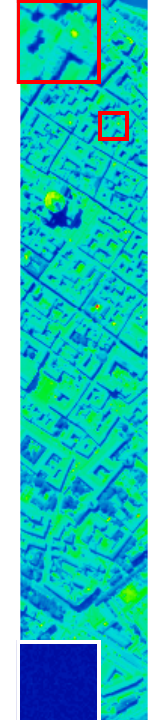}
		\\(c) $\text{SSRnet}$
		\\(\textit{45.11/\underline{2.98}})  
	\end{tabular}\hspace{-0.5cm}
	\begin{tabular}{c}	
		\includegraphics[ width=2cm,trim= 190 0 190 0,clip]{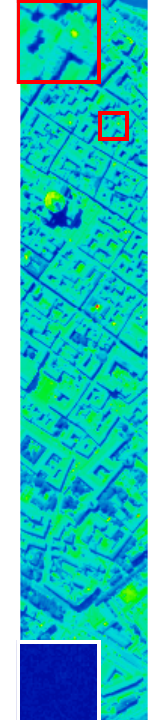}
		\\(d) $\text{MHFnet}$
		\\(\textit{42.82/4.37})  
	\end{tabular}\hspace{-0.5cm}
	
	\begin{tabular}{c}	
		\includegraphics[ width=2cm,trim= 190 0 190 0,clip]{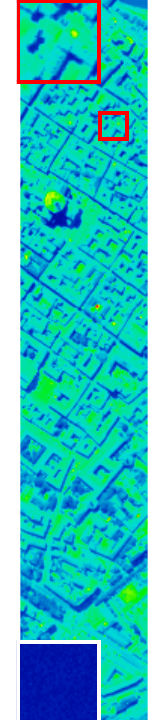}
		\\(e) $\text{EDBIN}$
		\\(\textit{\underline{45.35}/\textbf{2.93}})  
	\end{tabular}\hspace{-0.5cm}
	\begin{tabular}{c}	
		\includegraphics[ width=2cm,trim= 190 0 190 0,clip]{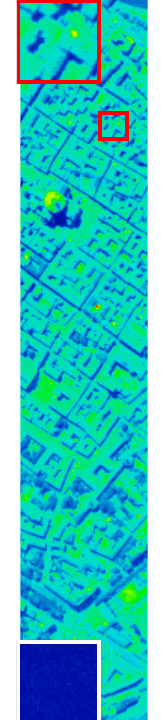}
		\\(f) $\text{DDPM-Fus}$
		\\(\textit{\textbf{45.39}/3.01})  
	\end{tabular}
	\vspace{0.2cm}
	\hspace{-0.6cm}
	\begin{tabular}{c}	
		\includegraphics[ width=2cm,trim= 190 0 190 0,clip]{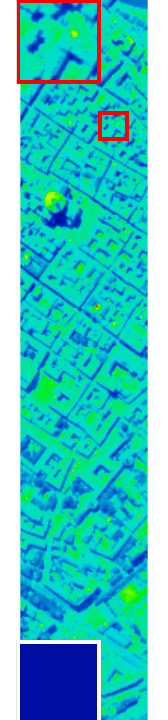}
		\\(g) $\text{GT}$
		\\(\textit{PSNR/SAM})  
	\end{tabular}\hspace{-0.5cm}
	\begin{tabular}{c}	
		\includegraphics[height=2.9cm,width=1.5cm]{fig/cycFusion/visualres/colorbar1}
	\end{tabular}

	\caption{(a-g) The 31st band (558 nm) of fused HrHSI (\textit{test image} in the Pavia Center dataset) obtained by the testing methods, where a ROI zoomed in 9 times (top-left) and the corresponding residual maps (bottom-left) are shown for detail visualization. PSNR and SAM are also listed for comparison.  
	}
	\label{fig.PaviaCenter}
\end{figure}

\begin{figure}[t]
	\centering
	\includegraphics[height=5cm,trim= 10 0 10 0,clip]{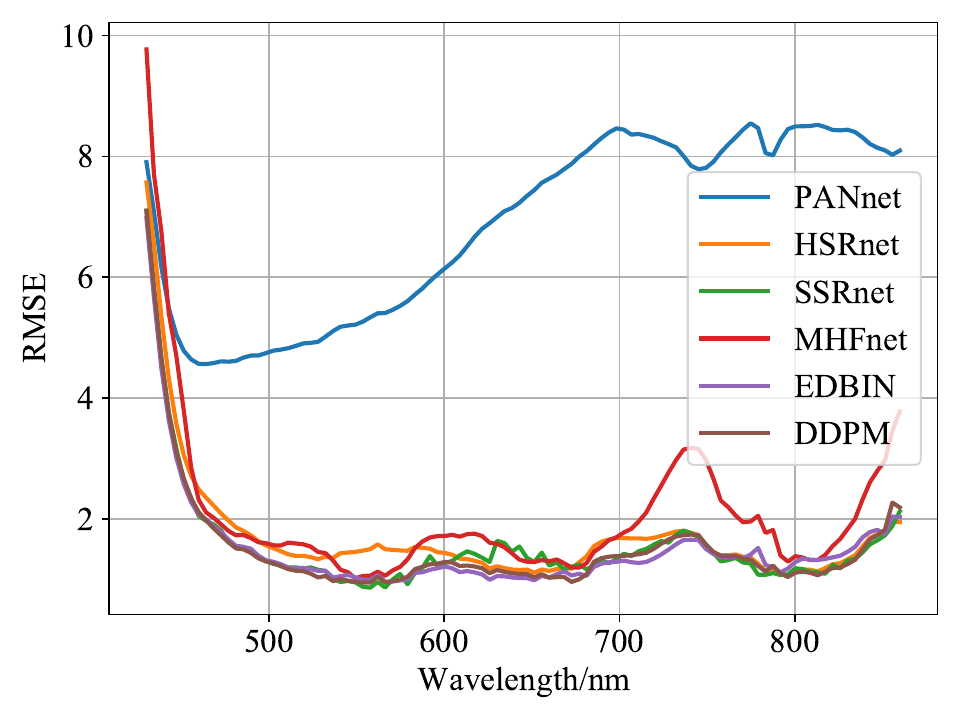}
	\caption{
		RMSE along with spectral bands for the comparison methods on the Pavia Center test data.
	}
	\label{fig.PaviaCenter.RMSE}
\end{figure}


\subsection{Experiments on the Remote Sensing Dataset}
We conduct two remote sensing datasets to further evaluate the fusion performance.
Remote sensing data, in contrast to indoor images, have lower spatial resolution and typically contain several spectral signatures in a single image.
The quantitative metrics of all testing methods over the Chikusei and Pavia Center datasets are shown in Table~\ref{tab.Chikusei}.
As can be seen, the SAM results tested on these two datasets are significantly smaller than the same metrics evaluated on the CAVE dataset in Table~\ref{tab.CAVE}.
Our proposed DDPM-Fus also obtains the best results in terms of the PSNR and is almost optimal in other indicators.
In addition, for visual comparison, we present one band in each fused image, as seen in Fig.~\ref{fig.Chikusei.region2} and Fig.~\ref{fig.PaviaCenter}.
Last, the corresponding RMSE in each band is also shown in Fig.~\ref{fig.Chikusei.region2} (h) and Fig.~\ref{fig.PaviaCenter.RMSE} and the multi-stage fusion models, such as SSRnet, EDBIN and ours DDPM-Fus are preferred to the one-stage fusion model, PANnet and HSRnet.

\subsection{Model Discussion}
\label{sec.modelDisscussion}
\subsubsection{Complexity Analysis}
The computational complexity of comparison models heavily depends on the number of parameters and convergence. 
We list the number of parameters of comparison models, the floating-point operations (FLOPs) when computing one image through each model once and corresponding training time for the three datasets is listed in Table~\ref{tab.complexity}. 
Benefiting from using the DDIM sampler, our proposed DDPM-Fus can achieve the performance of multi-stage fusion models by fusing only once.
Meanwhile, as can be seen in Table~\ref{tab.complexity}, FLOPs of our model and HSRnet are in the same order of magnitude.


\begin{table}[!t]
	\centering
	\renewcommand\arraystretch{1.2}
	\caption{Number of Parameters, FLOPs and training time of comparison models.}
		\label{tab.complexity}
	\begin{tabular}{cccc}
		\toprule[1.3pt]
		\multicolumn{4}{c}{CAVE}                                                                                                                                                 \\ \hline
		\multicolumn{1}{|c|}{}         & \multicolumn{1}{c|}{Size ($\times   10^{6}$)} & \multicolumn{1}{c|}{Flops ($\times   10^{9}$)} & \multicolumn{1}{c|}{Training time (h)} \\ \hline
		\multicolumn{1}{|c|}{PANnet}   & \multicolumn{1}{c|}{1.08}                      & \multicolumn{1}{c|}{4.41}                         &        \multicolumn{1}{c|}{3.26}      \\ \hline
		\multicolumn{1}{|c|}{HSRnet}   & \multicolumn{1}{c|}{10.24}                       & \multicolumn{1}{c|}{2.00}                          & \multicolumn{1}{c|}{2.38}                   \\ \hline
		\multicolumn{1}{|c|}{SSRnet}   & \multicolumn{1}{c|}{0.03}                       & \multicolumn{1}{c|}{0.11}                          & \multicolumn{1}{c|}{1.34}                    \\ \hline
		\multicolumn{1}{|c|}{MHFnet}   & \multicolumn{1}{c|}{2.16}                      & \multicolumn{1}{c|}{3.71}                          & \multicolumn{1}{c|}{5.94}                     \\ \hline
		\multicolumn{1}{|c|}{EDBIN}    & \multicolumn{1}{c|}{1.24}                      & \multicolumn{1}{c|}{13.34}                          & \multicolumn{1}{c|}{34.26}                     \\ \hline
		\multicolumn{1}{|c|}{DDPM-Fus} & \multicolumn{1}{c|}{1.69}                        & \multicolumn{1}{c|}{2.66}                          & \multicolumn{1}{c|}{11.44}                \\ \hline
		\multicolumn{4}{c}{Chikusei}                                                                                                                                             \\ \hline
		\multicolumn{1}{|c|}{}         & \multicolumn{1}{c|}{Size ($\times   10^{6}$)} & \multicolumn{1}{c|}{Flops ($\times   10^{9}$)} & \multicolumn{1}{c|}{Training time (h)} \\ \hline
		\multicolumn{1}{|c|}{PANnet}   & \multicolumn{1}{c|}{16.93}                         & \multicolumn{1}{c|}{69.33}                          & \multicolumn{1}{c|}{7.13}                  \\ \hline
		\multicolumn{1}{|c|}{HSRnet}   & \multicolumn{1}{c|}{156.45}                         & \multicolumn{1}{c|}{3.05}                          & \multicolumn{1}{c|}{13.15}                  \\ \hline
		\multicolumn{1}{|c|}{SSRnet}   & \multicolumn{1}{c|}{0.44}                         & \multicolumn{1}{c|}{1.81}                          & \multicolumn{1}{c|}{3.35}                  \\ \hline
		\multicolumn{1}{|c|}{MHFnet}   & \multicolumn{1}{c|}{29.30}                         & \multicolumn{1}{c|}{47.72}                          & \multicolumn{1}{c|}{7.67}                  \\ \hline
		\multicolumn{1}{|c|}{EDBIN}    & \multicolumn{1}{c|}{18.40}                         & \multicolumn{1}{c|}{23.19}                          & \multicolumn{1}{c|}{53.62}                  \\ \hline
		\multicolumn{1}{|c|}{DDPM-Fus} & \multicolumn{1}{c|}{1.93}                         & \multicolumn{1}{c|}{2.72}                          & \multicolumn{1}{c|}{11.91}                  \\ \hline
		\multicolumn{4}{c}{Pavia Center}                                                                                                                                         \\ \hline
		\multicolumn{1}{|c|}{}         & \multicolumn{1}{c|}{Size ($\times   10^{6}$)} & \multicolumn{1}{c|}{Flops ($\times   10^{9}$)} & \multicolumn{1}{c|}{Training time (h)} \\ \hline
		\multicolumn{1}{|c|}{PANnet}   & \multicolumn{1}{c|}{10.79}                         & \multicolumn{1}{c|}{44.18}                          & \multicolumn{1}{c|}{12.01}                  \\ \hline
		\multicolumn{1}{|c|}{HSRnet}   & \multicolumn{1}{c|}{100.33}                         & \multicolumn{1}{c|}{2.70}                          & \multicolumn{1}{c|}{19.55}                  \\ \hline
		\multicolumn{1}{|c|}{SSRnet}   & \multicolumn{1}{c|}{0.28}                         & \multicolumn{1}{c|}{1.15}                          & \multicolumn{1}{c|}{6.88}                  \\ \hline
		\multicolumn{1}{|c|}{MHFnet}   & \multicolumn{1}{c|}{19.01}                         & \multicolumn{1}{c|}{31.00}                          & \multicolumn{1}{c|}{8.37}                  \\ \hline
		\multicolumn{1}{|c|}{EDBIN}    & \multicolumn{1}{c|}{11.76}                         & \multicolumn{1}{c|}{19.88}                          & \multicolumn{1}{c|}{45.45}                  \\ \hline
		\multicolumn{1}{|c|}{DDPM-Fus} & \multicolumn{1}{c|}{1.86}                         & \multicolumn{1}{c|}{2.70}                          & \multicolumn{1}{c|}{12.95}                  \\ \hline
			\bottomrule[1.3pt] 
	\end{tabular}
\end{table}
\subsubsection{Ablation Study}
We also perform the fusion task with different loss functions and sampling steps to investigate the optimal setting.
We implement an experiment conducted on the CAVE dataset and PSNR metrics are shown in Table~\ref{tab.ablation}. 
The results show that the $\ell_1$ loss is superior to $\ell_2$. Besides, the fusion performance is better when
the number of sampling steps in the test phase  is set to 1, 2 or 5.  Considering the efficiency of the execution, we set the sampling step to 1.

\begin{table}[t]
	\centering
	\renewcommand\arraystretch{1.2}
	\caption{
	PSNR and fusion time for fusing per image with different sampling steps and ablation study on loss functions  conducted on the CAVE dataset.}
	\label{tab.ablation}
	\begin{tabular}{c|c|c|c|c|c|c}
		\toprule[1.3pt]
		Sampling steps  & 50 & 20 & 10 & 5 & 2 & 1 \\ \hline
		$\ell_1$                  &  42.82   &43.54    & 43.68   & 43.68 & 43.66  &   43.66  \\ \hline
		$\ell_2$                  &  37.29  &  37.49  &  37.62  & 37.73  &  37.75 & 37.61  \\ \hline	
		Test time (s)          &  21.25  &  9.33  &  5.42  &3.75  & 2.5 & 1.92  \\ 
		\bottomrule[1.3pt]  
	\end{tabular}
\end{table}
\section{Conclusion}
\label{sec.sec5}
In this article, we proposed a novel
supervised HSI-MSI fusion model based on the conditional denoising diffusion probabilistic model, namely DDPM-Fus.
The DDPM-Fus exploit the spatial details and spectral characteristics in the HrHSI via learning the conditional generative model.
A U-net is used to learn the noise added in the forward diffusion process.
After the DDPM-Fus is trained, the desired HrHSI in the test data can be generated by performing the conditional noise reduction through the reverse process step-by-step.
The results of experiments conducted on three publicly available datasets demonstrated the effectiveness and efficiency of our proposed DDPM-Fus.
In future research, we will introduce 
the pre-trained generative diffusion model to enhance the quality of the fused image
and 
develop faster sampling models to reduce fusion time to further improve model efficiency.


\appendices
\section*{Appendix}
In this section, we present the derivation of the evidence lower bound (ELBO) \eqref{eq.elboKL}.

In the DDPM model, all corrupted images, $\mathbf{X}_1,\mathbf{X}_2,\dots,\mathbf{X}_T$, are seen as latent matrices.
From the point of variational inference,
the true posterior distribution $p(\mathbf{X}_{1:T}|\mathbf{X}_0)$ is approximated by a variational distribution $q(\mathbf{X}_{1:T} |\mathbf{X}_{0})$, where $\mathbf{X}_{1:T}$ represents  $\mathbf{X}_1,\mathbf{X}_2,\dots,\mathbf{X}_T$ for short.
Generally, we minimize the KL divergence between these two distributions as
\begin{align}
	&\text{KL}(q(\mathbf{X}_{1:T}|\mathbf{X}_{0})||p(\mathbf{X}_{1:T}|\mathbf{X}_0))\nonumber\\
	=&\int q(\mathbf{X}_{1:T}|\mathbf{X}_{0}) \log \frac{q(\mathbf{X}_{1:T}|\mathbf{X}_{0})}{p(\mathbf{X}_{1:T}|\mathbf{X}_0)}d\mathbf{X}_{1:T}\nonumber\\
	=&\int q(\mathbf{X}_{1:T}|\mathbf{X}_{0}) \log \frac{q(\mathbf{X}_{1:T}|\mathbf{X}_{0})p(\mathbf{X}_{0})}{p(\mathbf{X}_{0:T})}d\mathbf{X}_{1:T}\nonumber\\
	=&\int q(\mathbf{X}_{1:T}|\mathbf{X}_{0}) \log \frac{q(\mathbf{X}_{1:T}|\mathbf{X}_{0})}{p(\mathbf{X}_{0:T})}d\mathbf{X}_{1:T}+\log p(\mathbf{X}_{0})\nonumber\\
	=& \log p(\mathbf{X}_{0})-\mathbb{E}_q \left [\log \frac{p(\mathbf{X}_{0:T})}{q(\mathbf{X}_{1:T}|\mathbf{X}_{0})}\right ]
\end{align}
Due to the nonnegativity of KL divergence,
we get the ELBO of the log-likelihood as
\begin{align}
\mathbb{E}_q \left [\log \frac{p(\mathbf{X}_{0:T})}{q(\mathbf{X}_{1:T}|\mathbf{X}_{0})}\right ] \leq \log p(\mathbf{X}_{0}).
\end{align}
Therefore we can maximize the ELBO to achieve the maximum likelihood estimate.
Note that the joint distribution of all variables in the forward diffusion and reverse diffusion  are
\begin{align}
	q(\mathbf{X}_1,\mathbf{X}_2,\cdots,\mathbf{X}_T|\mathbf{X}_0)&=\prod^{T}_{t=1} q(\mathbf{X}_t|\mathbf{X}_{t-1}),\\
	 p(\mathbf{X}_0,\mathbf{X}_1,\mathbf{X}_2,\cdots,\mathbf{X}_T)&=p(\mathbf{X}_{T})\prod^{T}_{t=1} p_\theta(\mathbf{X}_{t-1}|\mathbf{X}_{t}).
\end{align}
Then, the ELBO can be further simplified as 
\begin{align}
	&\mathbb{E}_q \left [\log \frac{p(\mathbf{X}_{0:T})}{q(\mathbf{X}_{1:T}|\mathbf{X}_{0})}\right ]\nonumber\\
	=&\mathbb{E}_q \left [   \log p(\mathbf{X}_{T})+\log \frac{p(\mathbf{X}_{0:T-1})}{q(\mathbf{X}_{1:T}|\mathbf{X}_{0})}\right ]\nonumber\\
	=& \mathbb{E}_q \left [   \log p(\mathbf{X}_{T})+\sum\limits _{t=1}^T\log   \frac{p_\theta(\mathbf{X}_{t-1}|\mathbf{X}_{t})}{q(\mathbf{X}_{t}|\mathbf{X}_{t-1})}\right ]\nonumber\\
	=&\mathbb{E}_q \left [   \log p(\mathbf{X}_{T})+\sum\limits _{t=1}^T\log   \frac{p_\theta(\mathbf{X}_{t-1}|\mathbf{X}_{t})}{q(\mathbf{X}_{t-1}|\mathbf{X}_{t},\mathbf{X}_{0})}\cdot\frac{q(\mathbf{X}_{t-1}| \mathbf{X}_{0})}{q( \mathbf{X}_{t}|\mathbf{X}_{0})}\right ]\nonumber\\
	=& \mathbb{E}_q \left [   \log\frac{p(\mathbf{X}_{T})}{q( \mathbf{X}_{T}|\mathbf{X}_{0})} +\sum\limits _{t=1}^T\log   \frac{p_\theta(\mathbf{X}_{t-1}|\mathbf{X}_{t})}{q(\mathbf{X}_{t-1}|\mathbf{X}_{t},\mathbf{X}_{0})}\right ]\nonumber\\
	=& -\text{KL}\left(q(\mathbf{X}_{T}|\mathbf{X}_{0})||p(\mathbf{X}_{T} )\right) \nonumber\\
	&-\sum_{t=1}^T \text{KL}\left(q(\mathbf{X}_{t-1}|\mathbf{X}_{t},\mathbf{X}_{0})||p_\theta(\mathbf{X}_{t-1}|\mathbf{X}_{t})\right).
\end{align}
Note that the first term is 
KL divergence between the distribution of $\mathbf{X}_T$, the output of the forward diffusion process, and the prior distribution $p(\mathbf{X}_{T})$, which is very close to 0 due to the Gaussian diffusion kernel used in the forward process.
We ignore the first term, thus, the second term is the desired result  \eqref{eq.elboKL}.

 
%

\bibliographystyle{bib/IEEEtran.bst}
\bibliography{bib/strings.bib}

\newpage

\vfill

\end{document}